\documentclass[preprint]{aastex}
\usepackage{color,wrapfig}

\newcommand{\WF}{{\em WFIRST}}
\newcommand{\Spitzer}{{\em Spitzer}}
\newcommand{\Kepler}{{\em Kepler}}
\newcommand{\TESS}{{\em TESS}}
\newcommand{\HST}{{\em HST}}
\newcommand{\Euclid}{{\em Euclid}}
\newcommand{\GAIA}{{\em Gaia}}
\newcommand{\piE}{\pi_{\rm E}}
\newcommand{\thetaE}{\theta_{\rm E}}
\newcommand{\tE}{t_{\rm E}}

\begin{document}

\begin{titlepage}
\begin{center}

{\huge\bf NASA ExoPAG Study Analysis Group 11:

Preparing for the 

WFIRST Microlensing Survey}

\vspace{1in}

{\Large Jennifer C. Yee$^1$ (chair), 
Michael Albrow$^2$,
Richard K. Barry$^3$,
David Bennett$^4$,
Geoff Bryden$^5$,
Sun-Ju Chung$^6$,
B. Scott Gaudi$^7$,
Neil Gehrels$^3$,
Andrew Gould$^7$,
Matthew T. Penny$^{7}$,
Nicholas Rattenbury$^8$,
Yoon-Hyun Ryu$^{6, 9}$,
Jan Skowron$^{10}$,
Rachel Street$^{11}$,
Takahiro Sumi$^{12}$
}
\end{center}

\vfill

{\small
\noindent
$^1$ Sagan Fellow, Harvard-Smithsonian Center for Astrophysics, 60 Garden St., Cambridge, MA 02138, USA\\
$^2$ Department of Physics and Astronomy, University of Canterbury, Private Bag 4800, Christchurch, New Zealand\\
$^3$ Astrophysics Science Division, NASA Goddard Space Flight Center, Greenbelt, MD 20771, USA\\
$^4$ University of Notre Dame, Department of Physics, 225 Nieuwland Science Hall, Notre Dame, IN 46556-5670, USA\\
$^5$ Jet Propulsion Laboratory, M/S 169-506, 4800 Oak Grove Drive, Pasadena, CA, 91109, USA\\
$^6$ Korea Astronomy and Space Science Institute, Hwaam-Dong, Yuseong-Gu, Daejeon 305-348, Republic of Korea\\
$^7$ Department of Astronomy, Ohio State University, 140 West 18th Avenue, Columbus, OH 43210, USA\\
$^8$ Department of Physics, University of Auckland, Private Bag 92-019, Auckland 1001, New Zealand\\
$^9$ Department of Astronomy and Atmospheric Sciences, Kyungpook National University, Daegu 702-701, Republic of Korea\\
$^{10}$ Warsaw University Observatory, Al. Ujazdowskie 4, 00-478 Warszawa, Poland\\
$^{11}$ Las Cumbres Observatory Global Telescope Network, 6740 Cortona Drive, Suite 102, Goleta, CA 93117, USA\\
$^{12}$ Department of Earth and Space Science, Osaka University, Osaka 560-0043, Japan\\
}

\begin{center}

{\Large\today}

\end{center}
\end{titlepage}

\section*{Executive Summary}
\addcontentsline{toc}{section}{Executive Summary}

The \WF\, microlensing survey will observe tens of thousands of
microlensing events and detect thousands of planets, including
free-floating planets. Because of its high resolution and high
photometric accuracy, \WF\, will characterize these
events at an unprecedented level of precision. It will be able to
routinely observe higher-order microlensing effects, which can be
combined in various ways to yield masses for the lens (host) stars and
therefore the true masses of the planets (rather than just the mass
ratio $q=m_p/M_{\rm star}$). In addition, because this survey will be performed in the near-IR,
\WF\, will be able to observe extincted parts of the Galactic bulge
not covered by optical microlensing surveys, and
  because it is in space, \WF\, will reach magnitudes far fainter than
  can be achieved from the ground. However, because \WF\, will far exceed the
capabilities of ground-based microlensing surveys, the exact potential
of these new opportunities is unknown. {\bf We identify the following
  programs that will enhance \WF\, microlensing science and reduce the
  mission's scientific risk.}

\subsection*{1. Precursor \HST\, Observations}

The \WF\, microlensing survey will be the first survey capable of
systematically measuring the astrometric microlensing effect, which
gives a measurement of the angular size of the Einstein ring,
$\thetaE$, and the relative proper motion between the source and the
lens. Measuring these effects is vital for measuring lens masses with
\WF. They can be combined with other effects to measure masses even
for planets with faint host stars (e.g., brown dwarfs) and for planets
detected without caustic crossings. However, most astrometric
microlensing signals will be close to the
signal-to-noise ratio limit, and because this effect is so
difficult to measure without \WF-like capabilities, the systematics
are unknown.

{\bf The best way to validate the \WF\, astrometric microlensing
  measurements is to conduct precursor \HST\, imaging
  of the \WF\, field $\sim 10$ years in advance.} Optical,
single-epoch, imaging of the field at this time will
resolve 20\% of the lenses and sources that will be future \WF\,
microlensing events. The observed \HST\, separation gives a direct
measurement of the source-lens relative proper motion for comparison
with the astrometric microlensing results. In addition, these proper
motion measurements can be used to calculate $\thetaE$ for events
without observable finite source or astrometric
microlensing effects.

In addition, {\bf multi-epoch observations of a few fields using
  \HST/WFC3/IR will provide a valuable data set for testing the \WF\,
  photometry/astrometry pipeline, the development of which is
  mission-critical}.

\subsection*{2. A Ground-Based, Near-IR, Microlensing Survey}

The near-IR microlensing event rate in the potential \WF\, fields has
never been measured. Much of our current understanding of these fields
comes from extrapolations of optical microlensing surveys both to
redder wavelengths and lower Galactic latitude. Improving our
understanding of these fields is crucial both for reducing scientific
risk from uncertainties in the simulations of the \WF\, mission and
for optimizing the microlensing field to maximize \WF's scientific
return. {\bf The best way to characterize the \WF\, field is to
  conduct a near-IR microlensing survey from the ground.} This will
provide a direct measurement of the microlensing event rate for bright
stars and a preliminary understanding of the source luminosity
function. Thus, it will reduce scientific risk for \WF\, by improving
our simulations and predictions for the \WF\, planet yield.

\subsection*{3. Development of Expertise}

There are also several other programs that will support \WF\,
science. {\bf Projects to measure parallax (e.g., with \Spitzer\, or
  \Kepler), astrometric microlensing, and lens fluxes
  with current instrumentation will all help to develop the techniques
  that \WF\, will use to measure lens masses while providing
  preliminary data on the relative populations of planets in the disk
  and the bulge.} In addition, a competition in microlensing analysis
could stimulate the development of new techniques to handle the vast
microlensing data set that \WF\, will produce.

\pagebreak

\tableofcontents

\pagebreak

\section{Introduction}
\label{sec:intro}

The Wide-Field Infrared Survey Telescope (\WF)
microlensing survey will be the most powerful microlensing survey ever
undertaken, detecting an order of magnitude more microlensing events
than available from the ground, and characterizing them with much
higher precision. As a result, it will detect thousands of exoplanets
ranging from a few lunar masses to super-Jupiters. More details about
the \WF\, exoplanet survey can be found in the
\WF-AFTA Science Definition Team Final Report
  \citep{Spergel13}. Because \WF\, microlensing represents a major
advance in microlensing capability, it will be able to reach new areas
of the bulge and regularly observe microlensing
effects rarely measurable in ground-based
observations. As a result, unlike all previous microlensing surveys,
\WF\, will be able to routinely measure the masses of many of its
planets and their host stars.

The purpose of this SAG is to consider ways of
reducing the technical risk of the \WF\, microlensing
  survey, as well as maximizing and expanding its science
  potential. We have identified several 
  programs that are critical for enabling the following three key areas:

\fbox{\parbox[c][][c]{0.9\textwidth}{\bf
\begin{enumerate}
\item{Absolute planet mass measurements with \WF,}
\item{\WF\, field selection and simulations,}
\item{Development of microlensing expertise.}
\end{enumerate}
}}

\noindent These topics will be discussed in further detail
below. Sections \ref{sec:HST}--\ref{sec:chall} will then discuss
specific programs that can address these points. The first program,
described in Section \ref{sec:HST}, is optical {\it Hubble Space
  Telescope\,}(\HST) imaging of the entire \WF\, field that can be
used to validate \WF\, astrometric microlensing
measurements, which will be used to measure planet masses. These
\HST\, data can also be used to measure the source luminosity function
in the bulge. The second program is a ground-based, near-IR
microlensing survey (Section \ref{sec:hband}) that will provide
concrete data on the expected near-IR microlensing event rates in the
\WF\, field, reducing our reliance on extrapolation and Galactic
models for predicting \WF\, yields and finalizing field
selection. Sections \ref{sec:parallax}--\ref{sec:hstast} discuss
observational programs to measure microlens parallax, lens fluxes, and
astrometric microlensing of current microlensing
events, supporting the continued development of techniques \WF\, will
use to measure lens and planet masses.  Sections \ref{sec:HST-IR} and
\ref{sec:chall} discuss programs that can help us meet the challenge
of analyzing the \WF\, microlensing data. Our conclusions are given in
Section \ref{sec:conclusions}. The appendices provide background
information and explore topics outside of the scope of the main
report.

\subsection{\WF\, Planet Masses}

\fbox{\parbox[c][][c]{0.9\textwidth}{\bf The most important need is to
    maximize the number of lens mass and thus planet
      mass measurements that can be made for \WF\, microlensing
    events. }}

Aside from the sheer number of microlensing events, a key advantage of
\WF\, over ground-based microlensing surveys is its ability to
routinely measure lens masses (Appendix \ref{app:intro}). While \WF\,
microlensing will robustly measures {\it mass ratios} ($q=m_p/M_{\rm
  L}$) for exoplanets, accurately determining the {\it masses} of
those planets requires accurately measuring the masses of the lens
(host) stars. Because of the higher resolution and better photometric
accuracy available from space, \WF\, can take advantage of a variety
of techniques for measuring or inferring the masses of the lenses that
are rarely available from the ground
\citep{BennettRhie02,Bennett07}. In many cases \WF\,
will measure lens masses from measurements of the lens flux, microlens
parallax \citep{Gould13,Yee13}, or measurements of astrometric
microlensing \citep{GouldYee14}. However, while the theory of these
various techniques for measuring masses is well understood, to date,
they have only been practically applied in a handful of cases
\citep[e.g., ][]{Bennett06,Gaudi08,Janczak10}.  It is
vital that we continue to develop expertise in these areas in
preparation for the \WF\, mission. We must also be able to vet these
measurements to confirm their accuracy in the face of unknown
systematics.

If the true planet masses are measured for a large fraction of \WF\,
events, they can reveal detailed structure in the planetary mass
distribution. In addition, because a measurement of the lens mass
gives a measurement of its distance from the Earth,
true planet masses also mean measurements of the planet frequency as a
function of Galactic environment. Furthermore, these mass measurements
can be used to identify microlensing events caused by old brown
dwarfs, stellar-mass black holes, and neutron stars, objects that
cannot otherwise be found unless they have luminous companions.

\begin{figure}[t!]
\fbox{
\begin{minipage}{0.6\textwidth}
  \includegraphics[width=0.95\textwidth]{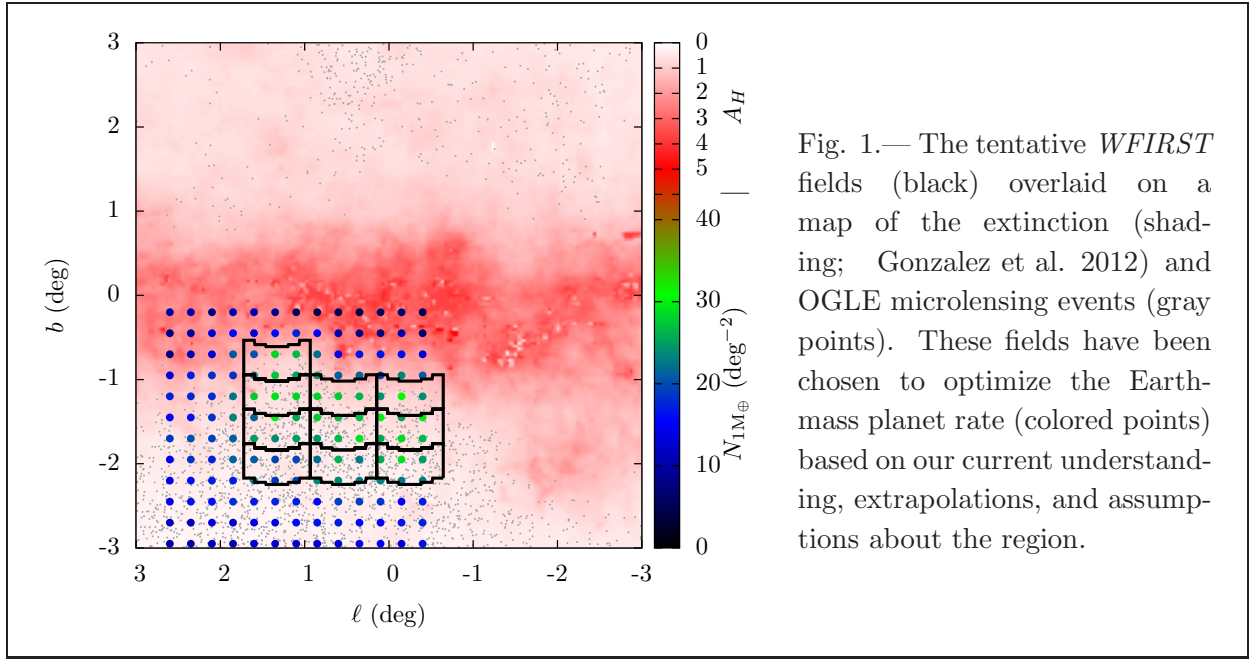}
\end{minipage}
\begin{minipage}{0.333\textwidth}
  \caption{The tentative \WF\, fields (black) overlaid
      on a map of the extinction \citep[shading; ][]{Gonzalezetal12}
    and OGLE microlensing events (gray points). These fields have been
    chosen to optimize the Earth-mass planet rate (colored points)
    based on our current understanding, extrapolations, and
    assumptions about the region.}
  \label{fig:wffield}
\end{minipage}
}
\end{figure}

\subsection{Field Selection}
\label{sec:intro-fields}

\fbox{\parbox[c][][c]{0.85\textwidth}{\bf The second need is to improve
    characterization of the \WF\, fields.}}

A number of assumptions and extrapolations are used in the
microlensing simulations for \WF. Currently, we rely either on
theoretical models of the Galaxy, such as the Besan\c{c}on model
\citep{Robinetal03}, or on the source luminosity function measured
from Baade's window combined with microlensing event rates
extrapolated from optical microlensing fields. There are already
suggestions of conflict between the measured optical, microlensing
event rates and those predicted by Galactic models \citep{Sumietal13},
which implies more severe uncertainties going redward in wavelength
into the near-IR and going spatially in toward the Galactic
center. 

\pagebreak
{\bf The \WF-AFTA report \citep{Spergel13} specifically cites the
  need to accurately measure
\begin{itemize}
\item{the source luminosity function,}
\item{the near-IR event rate,}
\item{the relative bulge-to-disk planet frequency.}
\end{itemize}} 

The exact placement of the \WF\, fields has not been
finalized. Improving these assumptions or replacing them with data is
necessary for accurately predicting the planetary yields and optimally
selecting the \WF\, fields.

\subsection{Microlensing Techniques}

\fbox{\parbox[c][][c]{0.8\textwidth}{\bf The third need is continued
    development of microlensing techniques.}}

The \WF\, microlensing mission will produce an enormous data set whose
analysis will be a massive undertaking. Moreover,
\WF\, will routinely measure higher-order light curve effects
(Appendix \ref{app:intro}) that are rarely observed from the
ground. In preparation for the launch of \WF, {\it
  it is vital to develop human potential and experience with
  microlensing as well as the analysis tools that will be used for the
  \WF\, microlensing mission.} As such, we place particular emphasis
on programs that will develop the techniques important to \WF.

\section{Optical \HST\, Imaging of the \WF\, fields}
\label{sec:HST}

\fbox{\parbox[c][][c]{0.9\textwidth}{\bf Immediate optical \HST\,
    imaging of the \WF\, fields will allow proper motion measurements
    for a substantial fraction of \WF\, stars, which provides a direct
    test of \WF\, astrometric microlensing
    measurements. These measurements are vital for
    measuring the masses of planets with faint or non-luminous
    hosts.}}

\WF\, has several techniques at its disposal for measuring lens masses
(Appendix \ref{app:intro}), but the situation is fundamentally
different for luminous and non-luminous lenses. For luminous lenses,
direct detection of lens flux, combined with the image elongation and
color-dependent centroid shifts that this induces over the course of
the mission, yield lens mass measurements \citep{Spergel13}. For
non-luminous lenses, mass measurements require both measurements of
$\thetaE$ and $\piE$ from higher-order microlensing
effects. Astrometric microlensing is an important component of those
measurements and essential for events with planets detected without
caustic crossings (as well as isolated black holes and neutron
stars). The optical \HST\, observations proposed here are necessary to
understand the systematics in the astrometric techniques so that we
can extend \WF's ability to measure masses to non-luminous lenses.

\subsection{Astrometric Microlensing}

\begin{figure}[h!]
\fbox{
\begin{minipage}{0.55\textwidth}
  \includegraphics[width=0.95\textwidth]{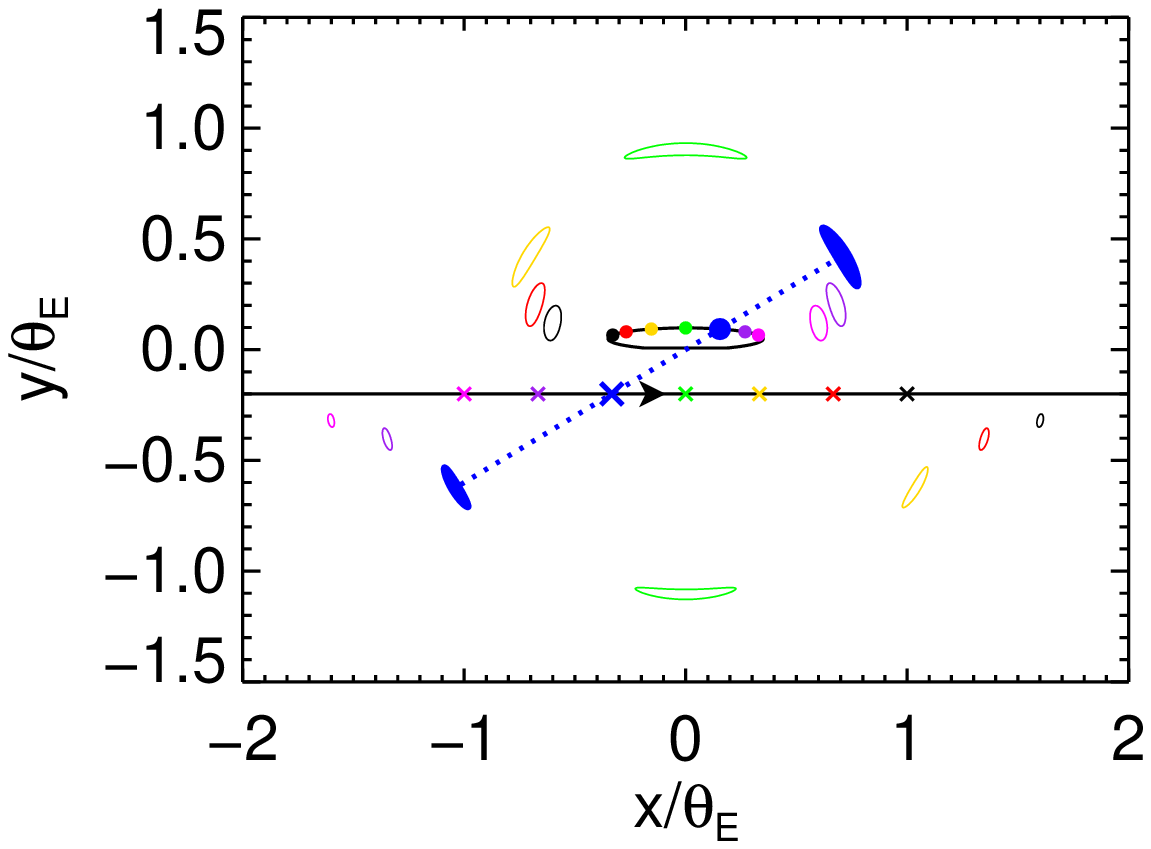}
\end{minipage}
\begin{minipage}{0.385\textwidth}
  \caption{In a microlensing event, the lens star passes
      in front of the source star creating two images (ovoids). The path
      of the lens is indicated by the arrow with specific positions
      marked by X's, color-coded to match the pairs of images
      created. The source star (not shown) is at the origin. Because the
      sizes and positions of the images are unequal, the apparent
      centroid of the light (filled circles) traces an ellipse on the
      sky creating an effect called ``astrometric microlensing''. {\bf
        Astrometric microlensing can be used to measure $\thetaE$ and
        ${\bf {\mu}}_{\rm rel}$}. \label{fig:ast}}
\end{minipage}
}
\end{figure}

Figure \ref{fig:ast} illustrates the astrometric microlensing
effect\footnote{``Astrometry,'' the measurement of the precise
  location of a target, appears in several different contexts
  throughout this report. In this section, we discuss {\it astrometric
    microlensing}, which is the effect of the lensing on the measured
  astrometry. The \HST\, observations proposed in this section are
  also astrometry, in that the key point is to measure the current
  positions of the future lenses and sources. In addition, the
  color-dependent centroid shifts described in Section
  \ref{sec:masses} and Appendix \ref{app:flux} are essentially
  astrometry. Finally, Section \ref{sec:HSTother} uses astrometry in
  the traditional sense to compare \WF\, measurements of the positions
  of stars {\it en masse} to what is possible with \GAIA.}  that arises
because the lensing of the source shifts the observed centroid of the
light. By measuring these small centroid shifts, we can measure both
the angular size of the Einstein ring, $\thetaE$, and the the
lens-source relative proper motion, {\bf$\mu_{\rm rel}$}. These
measurements can be combined with other information, one-dimensional
parallaxes in particular, to yield mass measurements for the lens
stars (see Appendix \ref{app:intro} for more details).  Because
astrometric microlensing only relies on detecting the light from the
source, it provides a crucial tool for measuring masses of faint or
dark lenses (e.g. brown dwarfs or black holes) and their
planets. Furthermore, because astrometric microlensing also measures
$\thetaE$, it supplies a critical piece of information necessary for
measuring masses for isolated (point lens) objects and two-body events
without caustic crossings (see Appendix \ref{app:intro}).

However, most astrometric microlensing measurements
will be close to or at the limit of the noise \citep{GouldYee14}. The
magnitude of the centroid shift due to astrometric microlensing,
$\Delta\theta_{\rm max}$, is given by:
\begin{equation}
\label{eqn:ast}
\Delta\theta_{\rm max} = \frac{\thetaE}{2\sqrt{u_0^2+2}} 
                 \sim 0.2 \mathrm{\,mas}
                          \sqrt{\frac{M_{\rm L}}{0.3 M_{\odot}}
                          \frac{\pi_{\rm rel}}{0.125 \mathrm{\,mas}}},
\end{equation}
where $M_{\rm L}$ is the lens mass and $u_0\lesssim 1$. To date, no measurements of this
effect have been published because they require high precision
astrometry, which will only be routinely available for microlensing
events observed from space.  Hence, it is vitally important to have
robust proper motion measurements for a substantial
number of \WF\, events to validate this technique. This is only
possible by comparison to \HST\, imaging of the \WF\, field made in
the optical, 10 years in advance of the \WF\, microlensing mission.

\subsection{Program Description}

Optical, single-epoch, imaging with \HST\, that
covers the entire \WF\, field would vastly increase the value of the
\WF\, microlensing survey. The primary goal of such
imaging would be to separately resolve the future
source and lens stars. The current lens-source
separation, divided by the time baseline between the \HST\,
observations and the observed microlensing event,
directly yields a measurement of their relative proper motion. This
need to measure proper motions for a substantial fraction of \WF\,
events drives the design of the program.

First, it is critical that the observations be undertaken
immediately. In observations taken now (10 years
before the \WF\, mission), 22\% of sources and lenses
are separated by at least 80 mas (i.e., are resolvable), but that
number declines rapidly as we get closer to the \WF\, launch date
($N \propto \exp{(-\Delta t)^2} \rightarrow$ 16\% and
9\% with a time baseline of 9 and 8 years, respectively). Second, the
observations must be done in the optical, where the diffraction limit
is lower, in order to resolve as many source-lens pairs as
possible. Finally, we note that previous observations of the bulge are
insufficient to meet the goals of this program. To
date, \HST\, coverage of the inner bulge has been sparse. There are
only 5 sets of \HST\, fields within the proposed \WF\, survey
area. These cover $<5$\% of the \WF\, field and are heavily biased to
the lower extinction, higher $|b|$ part of the field. Hence, to
measure proper motions of a significant fraction of \WF\, events
requires new observations.

Because of the size of the \WF\, field, imaging the entire field in
just one optical band will require $\sim750$ pointings of \HST. While
this number is large, the program could be split into several stages
to be undertaken over multiple cycles. To begin with,
sparse, preliminary observations in both F814W($I$) and F555W($V$) can
be used to assess the necessary depth and extent of the imaging
(especially the utility of F555W data in highly extincted
fields). These initial fields will focus on those
  guaranteed to be part of the final \WF\, field, since while the
  field may undergo some optimization, it will almost certainly have
  significant overlap with what is proposed now. These initial \HST\,
  fields can be followed by a larger program to capture a larger
area, and additional, targeted observations for regions where previous
data show it is beneficial to go deeper. In addition, while optical
data are being taken with ACS, complementary data in F160W($H$) or
F105W($Y$) can be taken simultaneously (see Section \ref{sec:HST-IR}).

\subsection{Additional \WF\, Microlensing Science}

Optical \HST\, imaging has several benefits in addition to measuring
relative proper motions for a large subset of \WF\, events.  First,
the improved accuracy of the positions and proper motions of the stars
from the optical data will also improve the data reduction of the
fields. Second, direct detections of the flux from the lens stars will
be another means for \WF\, to estimate lens masses in cases where the
lens is luminous. These optical data will provide additional colors
that will improve characterization of these lenses.  Also, the
luminosity functions derived from these data will improve our
understanding of the stellar populations in these fields, leading to
better simulations of the \WF\, microlensing survey and improving
estimates of the event rate and planet yields. Finally, \HST\,
relative proper motion measurements can give measurements of $\thetaE$
(Equation \ref{eqn:te}) for events without finite source or
astrometric microlensing measurements; $\thetaE$ is a
crucial parameter for microlensing planetary mass measurements.

Taken to the extreme, a 2-epoch, proper motion survey with \HST\, could
be used to search for stellar streams that provide more favorable
conditions for detecting planets in the habitable zone (see Appendix
\ref{app:hz}). The observed distribution of proper motions can also be
used to improve Galactic models and therefore predictions for
microlensing event rates.

\subsection{Other Science}
\label{sec:HSTother}

In addition to the direct benefits to the exoplanet mission, optical
\HST\, imaging can benefit the \WF\, mission in several other
ways. Most strikingly, the \WF\, microlensing mission
will provide a wealth of astrometric data on the Galactic bulge.
\GAIA\, will produce parallaxes for some of the stars in
  the bulge. However, the \WF\, relative parallaxes on these same
  stars will be $\ga 100$ times more precise. Furthermore, \WF\, will
  obtain relative parallaxes for millions of additional stars, below
  the \GAIA\, magnitude limit, with a precision of $\sigma(\pi)<4\,\mu$as
  for 40 million stars and $\sigma(\pi)<10\,\mu$as for an additional
  120 million stars.  \WF\, relative astrometry can be transformed to the
absolute \GAIA\, astrometric system with a precision of $\ll 1 \mu$as. When
the \WF\, data are combined with optical data, these additional colors
can be used to disentangle temperature, extinction, and metallicity of
stars and hence, measure detailed structure of the Galaxy including
the structure of the Galactic bar and spiral arms, as well as obtain
stellar age distributions along the line-of-sight from the Sun to the
Galactic center.  Finally, where the \HST\, imaging overlaps with
ground-based microlensing fields, the data can be used to measure lens
fluxes (and hence, masses) for events discovered from the ground (see
Section \ref{sec:masses}).

\section{Ground-based, Near-IR, Microlensing Survey}
\label{sec:hband}

\fbox{\parbox[c][][c]{0.9\textwidth}{\bf A ground-based,
    near-IR microlensing survey of the Galactic bulge
    would allow a direct measurement of the microlensing event rate in
    the \WF\, fields.  }}

\subsection{Unknown Event Rate in the \WF\, Field}

The \WF\, microlensing survey will be conducted in the near-IR. This
allows \WF\, to probe deeper into regions of the Galactic bulge that
have been inaccessible to optical surveys because of the high
extinction. Our understanding of microlensing in these fields and in
this band is limited to extrapolations from optical surveys in less
extincted regions. Hence, simulations of the \WF\, microlensing survey
are fundamentally limited by the lack of information about the
fields. 

Modeling of the inner Galaxy is an area of ongoing research. As such,
there are still many uncertainties and some discrepancies. For
example, there is significant dispersion among the predictions of the
microlensing event rates from various models. Also, the recent
measurement of the microlensing optical depth from the MOA survey,
which only partially overlaps with the \WF\, field, is discrepant by
as much as $2.8\sigma$ from the predictions of Galactic models
\citep{Sumietal13}. These problems could be even more severe for the
parts of the \WF\, field that are not covered by optical data.  For
instance, \citet{Sumietal13} estimate the microlensing event rate is
30-60\% higher than the values used in the original Science Definition
Team Final Report \citep{Green12}. The only way to resolve these
discrepancies is by obtaining more data. {\it In order to make good
  predictions for \WF\, microlensing, we need measurements of the
  microlensing event rate across the \WF\, field in the near-IR.}
Ultimately, these measurements may affect the ultimate field
placement.

\subsection{Description of the Survey}

We can measure the near-IR microlensing event rate from the ground for
bright microlensing events by conducting a microlensing survey in the
near-IR. This survey would cover all of the proposed \WF\, region,
areas deeper into the bulge, and overlap with the highest cadence
optical fields. It requires a good site in the southern hemisphere and
an IR camera with a large field-of-view. The ideal survey would be
dedicated to microlensing observations for the duration of the bulge
season (April-September) and achieve a cadence of 1 hr$^{-1}$, which
is necessary for a microlensing planet search. Based on these
requirements, the best existing facility for such a survey is the
VISTA telescope in Chile. Although the VVV program on the VISTA
telescope will nominally carry out a microlensing survey, the
observing cadence and duration is not actually optimized for
microlensing observations or planet detection and do not meet the
criteria specified above. Hence, a microlensing survey to measure the
near-IR microlensing event rate would best be performed as a separate
survey. Options beyond the VISTA telescope should be
  explored. For example, although its field-of-view is smaller and
  observing season shorter, a survey with UKIRT could achieve some of
  the major goals discussed here. Another alternative would be to
  build a wide-field IR camera and telescope such as the proposed
  WiFCOS camera coupled with the proposed Japanese telescope in
  Namibia.

\subsection{\WF\, Science Outcomes}

The primary purpose of the near-IR microlensing survey would be to
directly measure the event rate as a function of Galactic
coordinates. In regions of overlap with the optical, it will provide a
direct comparison of the microlensing event rates in the optical and
IR and quantify the relationship between them. This is important
because optical microlensing can reach much fainter events from the
ground as compared to observations in the IR.

In addition to event rates, the survey will provide measurements of
the near-IR source fluxes for the microlensing events. These
observations can be used to characterize the source
population. Together with the event rates, and possibly luminosity
functions from \HST\, (Section \ref{sec:HST}), these measurements can
be used to improve simulations of the \WF\, fields and probe the
uncertainties in the Galactic models. All of this will help us to
optimize \WF\, field selection.

Moreover, a near-IR microlensing survey can find giant planets in the
inner bulge where the extinction is too high for optical
surveys. These are only the tip of the iceberg of what
will be found by \WF. Incidentally, in the innermost
  regions of the bulge, the event timescale gives a strong
indication of the distance of the lens from the Galactic center
\citep{Gould95}. Hence, these planets can be used to estimate the
relative frequency of planets in the bulge and the disk without direct
knowledge of the lens mass. This is another unknown parameter that
affects field selection and contributes uncertainty to \WF\,
simulations.

Finally, we note that once such a survey is established, it could be
continued into the \WF\, era to yield additional microlensing parallax
measurements (see Appendix \ref{app:par}). Because the
  baseline between a geosynchronous orbit and the Earth is quite
  short, satellite parallax effects (Section \ref{sec:parallax}) will
  only be visible for the highest magnification events. Fortunately,
  this is the same subset that will be visible from the ground, where
  observations will be limited to the brightest (most highly
  magnified) targets because of the sky background. The situation is
more favorable if \WF\, is at L2 because of the longer
  baseline \citep{Yee13}.

\subsection{Other Science}

The immediate science benefit for ground-based microlensing from a
near-IR survey extends beyond the detection of additional planets in
the expanded survey area. The overlap with
  ground-based, optical fields will give near-IR data on optical
  microlensing events, which can provide important, additional
  information. Measurements of the source fluxes in the near-IR
improve the accuracy of lens flux measurements made with
high-resolution imaging, which is primarily done in the near-IR (see
Section \ref{sec:masses} and Appendix \ref{app:flux}). These lens flux
measurements lead to measurements of the masses and distances of the
lenses. Additionally, IR data can be used to check for chromatic
effects in the light curves such as lensing of starspots
\citep{Gouldetal13} or lensing of binary sources
\citep{Gaudi98,Hwangetal13}.

\section{Satellite Parallaxes}
\label{sec:parallax}

\fbox{\parbox[c][][c]{0.9\textwidth}{\bf Satellite observations of
    ground-based microlensing events will vastly increase the number
    of mass measurements for microlensing planets. The
      resulting distance measurements can be used to probe the
    relative frequency of planets in the bulge and the disk, which
    could affect \WF\, field selection.}}

\subsection{Microlensing Satellite Parallax}

In ground-based microlensing, it is difficult to
  determine the properties of the lens star, including its mass,
  unless microlens parallax is measured during the event. However, if
  this effect is measured, it is possible to measure the mass of the
  lens star and its planet.

\begin{figure}[ht]
\fbox{
\begin{minipage}{0.94\textwidth}
\includegraphics[width=\textwidth]{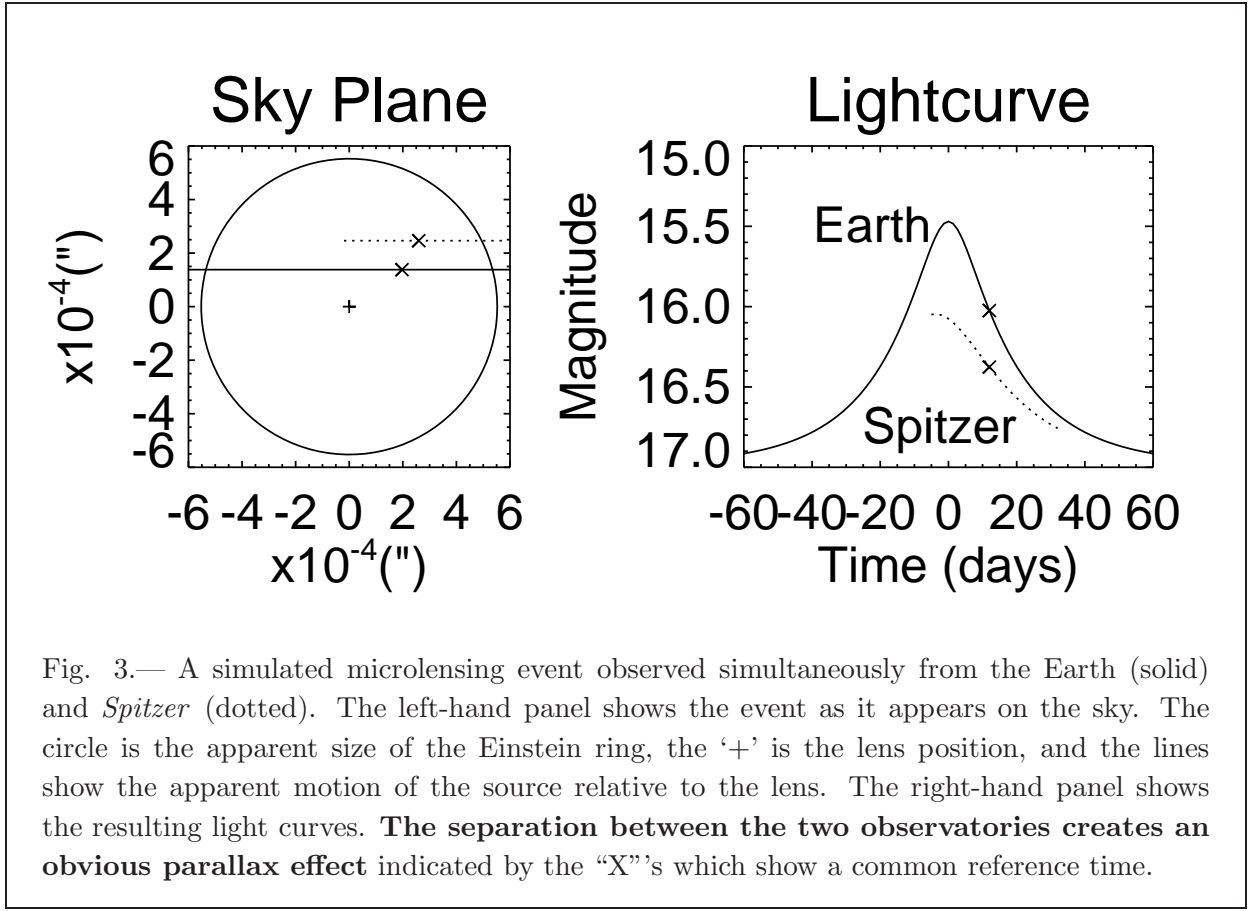}
\caption{A simulated microlensing event observed simultaneously from
  the Earth (solid) and \Spitzer\, (dotted). The left-hand panel shows
  the event as it appears on the sky. The circle is
    the apparent size of the Einstein ring, the `+' is the lens
    position, and the lines show the apparent motion of the source
    relative to the lens. The right-hand panel shows the resulting
  light curves. {\bf The separation between the two observatories
    creates an obvious parallax effect} indicated by the ``X'''s which
  show a common reference time.\label{fig:sat-lc}}
\end{minipage}
}
\end{figure}

Microlens parallax effects arise because microlensing
is a line-of-sight phenomenon. As such, observations of the same event
from two different locations, such as the Earth and a
  satellite, can yield very different light curves
due to parallax effects (Figure
\ref{fig:sat-lc}). The observable is $\piE=\pi_{\rm
  rel}/\thetaE$, the trigonometric parallax scaled to the size of the
Einstein ring. Appendix \ref{app:intro} shows that if
  both microlens parallax and $\thetaE$ are measured, this gives a
  measurement of the lens mass (Equation \ref{eqn:thetaE}).

Larger baselines lead to larger parallax effects, so satellites
well-separated from the Earth are the ideal platforms for these
observations. To date, only two such measurements have ever been
made. The first used the \Spitzer\, spacecraft to measure the parallax
of a binary lens in a microlensing event toward the
Small Magellanic Cloud \citep{Dong07}. The second was
a parallax measurement of a microlensing planet using the {\it Deep
  Impact} spacecraft \citep{Muraki11}.

Over 100 microlens parallax measurements per season are possible if
observations from a dedicated satellite at $\sim 1\,$ AU are combined
with ground-based observations.  Both \Spitzer\, and \Kepler\, are
well-suited for such dedicated campaigns to measure
this effect. In addition, \TESS\, may also be useful
for serendipitous measurements of microlens parallax.

\subsection{\Spitzer\, parallaxes}

\Spitzer\, can observe the Galactic bulge simultaneously with
ground-based observatories for 40 days a year. Due to its narrow
field-of-view, these observations would be targeted based on
identification of microlensing events from ground-based survey
data. Dedicating the satellite to microlensing observations during the
40 days (800 hours) will yield masses for 4--5 planets, including 1--2
planets discovered by the satellite, as well as microlens parallax
measurements for about 120 events. In addition to the exoplanet
science, this will identify brown-dwarf microlensing binaries and
provide a more accurate estimate of the stellar mass function
\citep{HanGould96}. A pilot program of 100 hours has been approved for
Cycle-10. However, only the full, 800-hour program is
  capable of independently detecting planets.

\subsection{\Kepler\, parallaxes}

A \Kepler\, microlensing mission complements a \Spitzer\, microlensing
mission. Because of its large field-of-view and predefined observing
program, the mission would target a specific subfield within the
larger \Kepler\, field-of-view. This field would be selected to have
the highest microlensing event rate as determined from ground-based
surveys. As such, the \Kepler\, mission will capture any microlensing
events that occur in this field during the observational campaign,
which can be identified afterwards from ground-based data, but it will
miss events outside of its field. Hence, the selection effects for
a \Kepler\, survey and a \Spitzer\, survey will be
substantially different, as will the actual events observed.

A \Kepler\, microlensing mission would enable mass measurements for
about 10 planets, half of which will be discovered by \Kepler\,. In
addition to achieving similar science goals to the \Spitzer\,
microlensing mission, because \Kepler\, is not targeted and captures
the whole field, it can improve our understanding of free-floating
planet candidates. Currently, the detection of free-floating planets
relies on statistical arguments that require a
  population of free-floating planets in addition to the stellar
  population in order to explain the excess of microlensing events
  with short timescales \citep{Sumietal11}. Through parallax
  measurements, \Kepler\, can identify whether specific free-floating
  planet candidates are actually due to stellar lenses or rule out
  that possibility. Furthermore, \Kepler\, is the perfect pilot for
a dedicated microlensing parallax satellite designed for
simultaneously observing all microlensing events for parallax
\citep{GouldHorne13}.

A microlensing mission is currently planned as part of {\it K2}. The
NGC 2158 observations taken in Campaign 0 will serve as a test data
set for developing a crowded-field {\it K2} reduction pipeline. Field
9 (to be observed in 2016) was specifically chosen to overlap with the
ground-based microlensing fields. Optimizing this field to cover the
microlensing fields with the highest event rate can influence the
expected yield by a factor of $\sim 1.5$ compared to fields $\sim 5$
degrees away. {\it A microlensing parallax survey with Field 9 offers
  the opportunity for unique science, especially since the proposed
  shutdown of \Spitzer\, would prevent the
  realization of the parallax survey proposed in the previous
  section.}

\subsection{\TESS\, parallaxes}
As part of its planned transit observations, \TESS\, will observe
portions of the Galactic bulge that overlap the
ground-based microlensing fields. Hence, it will
observe any microlensing events that occur within its fields during
normal observations, which could be analyzed for microlensing parallax
signals. Because the \TESS\, baseline is short ($\lesssim 10^{-3}\,$
AU) and the satellite is designed to observe the brightest stars
(albeit with much higher precision than necessary for microlensing),
parallax measurements will only be possible for the highest
magnification events or very bright events with caustic crossings
(e.g. planetary signals) that occur while \TESS\, is at
apoapse. However, if any such events do occur, they will be
identifiable from ground-based data, and all that will be required is
to analyze the serendipitous \TESS\, data.

If a bright, high-magnification event occurs during the \TESS\,
observations this offers the opportunity to measure not only
Earth-\TESS\, parallax but also a parallax signal from \TESS\, itself
as it executes its highly-elliptical orbit (Appendix
\ref{app:par}). This would be the first measurement of parallax from
satellite orbital motion. If \WF\, is in geosynchronous orbit, such
measurements will be the primary means to measure the second component
of the microlens parallax vector for events with peak
  magnification $\gtrsim 20$ \citep{Gould13}. This gives a complete parallax
  measurement, rather than a one-dimensional parallax, 
  which is necessary to measure planet masses. 

\subsection{Direct Relevance to \WF}

Satellite parallax observations of ground-based microlensing events
will yield planetary mass measurements as well as providing
opportunities for additional planet detections. These observations
will benefit \WF\, by contributing to our (currently nonexistent)
understanding of the relative frequency of planets in the bulge and
the disk. Furthermore, these will build expertise in parallax
measurements, which will be ubiquitous in the \WF\, data, albeit in a
somewhat different form (see Appendix \ref{app:par}). They can also
characterize the stellar mass function toward the bulge, which could
influence \WF\, simulations and field selection.

In addition, if \Euclid\, microlensing observations
  are ongoing at the time of the \WF\, microlensing survey, it is
  possible to measure satellite parallax effects between the two
  observatories. This offers an opportunity to directly measure masses
  for \WF\, free-floating planet candidates. Synergies between
  \Euclid\, and \WF\, are discussed in detail in Appendix
  \ref{sec:Euclid}.

Finally, \Kepler\, observations could serve as a pathfinder for a
small satellite mission to be flown simultaneously with \WF\, for the
purpose of measuring robust parallaxes for the large majority of \WF\,
events \citep{GouldHorne13}. The requirements for such a mission are
much less stringent than for the actual \WF\, mission since the events
and their positions as well as the existence of
planets can be identified from the \WF\, data. A
  dedicated parallax satellite can be positioned at $\sim 1$ AU from \WF, a
  better separation for parallax measurements than the Earth-L2
  baseline between \WF\, and \Euclid. \Kepler\, microlensing
observations can help to specify the exact requirements of such a
mission. For example, comparing the \Kepler\, and \Spitzer\,
microlensing events will clarify the trade-offs
entailed by larger pixels.

\section{Lens Flux Measurements of Current Microlensing Events}
\label{sec:masses}

\fbox{\parbox[c][][c]{0.9\textwidth}{\bf The mass of
      microlensing stars and planets can be directly measured if the
      light from the lens stars is measured with high-resolution
      imaging.}}

\subsection{Detecting Light from the Lenses}

Microlensing can detect planets around distant, low-luminosity
hosts. However, one of the main complications of ground-based
microlensing observations is that the Galactic bulge microlensing
fields have such a high stellar density that individual, main-sequence
stars are not resolved. Thus, there will often be excess light in the
seeing disk of the source in addition to the light from the source and the
lens. This makes it difficult to isolate the light from the lens star
 and prevents the lens, and its planet, from being
characterized.

The solution is to take high-resolution images sensitive enough to
detect low-mass stars (Appendix \ref{app:flux}). These observations
may be done while the lens and source are still superposed or after
waiting several years for the lens and the source to
separate. However, if the observations are taken while
  the lens and source are still superposed, in some cases there is a
  risk of ambiguous results. While the various probabilities can be
  calculated, there is a possibility that the excess flux is not due
  to the lens but due to a companion to the source or an extremely
  unlucky, chance superposition of an unrelated star
\citep{Janczak10}. This ambiguity can be resolved by
measuring the relative proper motion between the excess flux and the
source itself to ensure that the motion is consistent
with the predictions from the microlensing light curve. With
high-resolution images that reach the diffraction limit and are taken
a number of years after the event, it is possible to resolve the lens
and source stars, and therefore to verify the lens-source relative
proper motion measured from the planetary microlensing light curve
\citep[][in prep]{Batista14b}.

\subsection{Current Capabilities}

These observations may be taken from the ground using adaptive optics
with an 8-meter class telescope or, even better, from space using
\HST. Because AO observations must be done in the
near-IR, they require an estimate of the source flux in the
near-IR. If near-IR observations were not taken during the
microlensing event, this flux must be inferred from optical
data. \HST\, has the advantage that it can observe in the optical
where the source flux is always known. In addition, because the
diffraction limit is smaller, the source and the lens can be resolved
much sooner using \HST, enabling proper motion
measurements. Alternatively, the association between
  the excess flux and the lens can be confirmed by measuring the
  color-dependent centroid shift from \HST\, observations in multiple
  bands \citep{Bennett06}.

Among other things, previous high-resolution observations have
identified a massive, giant planet orbiting an M dwarf
\citep{Dong09_071}, a sub-Saturn mass planet likely to be in the
Galactic bulge \citep{Janczak10}, a cold super-Earth \citep{Kubas12},
and a Jupiter in the habitable zone \citep{Batista14}. Continued
observations of this type will characterize more microlensing lens
stars and lead to measurements of their planets' masses. Furthermore,
they will benefit \WF\, by building expertise in the technique and
increasing our understanding of planet occurrence as a function of
Galactic environment (Section \ref{sec:intro-fields}).

\section{Astrometric Microlensing Due to Black Holes}
\label{sec:hstast}

\fbox{\parbox[c][][c]{0.9\textwidth}{\bf Astrometric microlensing can
    be observed for microlensing events caused by black holes. Such
    observations will help develop the technique in advance of the
    \WF\, mission.}}

Equation \ref{eqn:ast} shows that astrometric microlensing
measurements are not usually possible without \WF\,
precision \citep[28 $\mu$as/observation;
  ][]{GouldYee14}. However, because the magnitude of the effect is
proportional to the square-root of the lens mass, for the subset of
events caused by stellar-mass black holes ($\sim 10 M_{\odot}$) the
signal increases to $\sim 1.1$ mas. In addition, candidate events can
be easily identified because the size of the Einstein ring increases
with increasing lens mass (Equation \ref{eqn:thetaE}), and
consequently, more massive objects have proportionally longer
timescales (Equation \ref{eqn:te}), typically hundreds of days for
black holes. Continued efforts to measure this effect using \HST\,
\citep[e.g.,][]{Sahu12} and adaptive optics will further the
development of this technique, which will be vital to the \WF\,
mission.

\section{Multi-Epoch, Near-IR, \HST\, Observations}
\label{sec:HST-IR}

\fbox{\parbox[c][][c]{0.9\textwidth}{\bf Multi-epoch, \HST/WFC3/IR
    observations of a few fields in the bulge
    are necessary for developing the \WF\, pipeline
    and understanding how well a random dither pattern will
    characterize the \WF\, detector.}}

Developing a robust photometry/astrometry pipeline for
  \WF\, is mission-critical. This process would benefit from test data
  with similar properties and systematics to the future \WF\,
  data. Such data do not currently exist. 

One major concern is that overlapping stars in the crowded fields
could induce systematics in the photometry and astrometry as the
amount of the overlap changes as a function of time due to proper
motion. A second concern is in understanding how well a random dither
pattern samples the \WF\, detector. If the detector is properly
sampled, {\it the \WF\, microlensing data will provide a multitude of
  point sources that can be used to understand the detector at a
  sub-pixel scale. Understanding the detector at this level is
  critical to the weak-lensing survey.} The same data needed to test
the photometry/astrometry pipeline can also be used to test the
usefulness of a random dither pattern in characterizing the \WF\,
detector.

Multi-epoch observations of the bulge with \HST/WFC3/IR would provide
the data to achieve both of these goals. Such a
program would require $\sim 10$ orbits per season for
a minimum of 3 seasons. This would give a time baseline of at least 2
years, enough to measure both parallax and proper motions for a select
number of fields and to test the \WF\, pipeline. Simultaneous optical
data will provide the higher resolution necessary to determine how
well the pipeline can reconstruct the underlying star pattern. In
addition, the IR data can be used to understand how
well a random dither pattern will characterize the detector using
crowded-field observations. Such a program could
overlap with the \HST\, observations described in
  Section \ref{sec:HST} or Section \ref{sec:masses}.

\section{Microlensing Analysis Challenge}
\label{sec:chall}

\fbox{\parbox[c][][c]{0.9\textwidth}{\bf An open competition in
    microlensing analysis will encourage the development of novel
    techniques and attract new people to the field.}}

\subsection{Complexities in Microlensing Analysis}

Many planetary microlensing events should be simple to
model. \citet{GouldLoeb92} and \citet{GaudiGould97} show that the mass
ratio and projected separation, $q$ and $s$, can be easily estimated
for a large subset of planetary, microlensing light curves. However,
this will not be universally true, especially since \WF's
unprecedented photometric accuracy and homogeneous data set will allow
detailed investigation and modeling of higher-order lensing effects
such as parallax and orbital motion. Furthermore, not
  all events will obey these simple relations. Of particular interest
are planets in binary star systems, \citep[i.e., triple lenses as in]
  []{Gouldetal14}, planets detectable without direct caustic crossings
  \citep[e.g., ][]{Zhu14}, and planets in high-magnification
  microlensing events \citep{GriestSafizadeh98}.

Although our techniques for solving microlensing light curves have
grown quite sophisticated \citep[e.g.,
][]{GouldGaucherel97,Dong06,Cassan08,Bennett10,Bozza10}, there still
remain unsolvable light curves.

\subsection{Parameters of the Challenge}

One way to address this problem would be to hold an open competition
in microlensing light curve analysis. This competition would be open
to anyone, and in particular should seek out the participation of
mathematicians and computer scientists. Bringing an outside
perspective to the problem and drawing on a breadth of expertise can
lead to the development of new algorithms and approaches to solving
the multivariate and multi-modal microlensing likelihood
space. Similar competitions have been carried out for weak
lensing\footnote{great3challenge.info} and strong
lensing\footnote{timedelaychallenge.org}.

The basic elements of the competition would be to:
\begin{enumerate}
\item{Provide a basic introduction to microlensing \citep[e.g., ][]{Gaudi12},}
\item{Provide a set of microlensing light curves (real or simulated or both),}
\item{Establish a metric with which to evaluate the results.}
\end{enumerate}

The goals of the competition would be to
\begin{itemize}
\item{Develop new analysis techniques,}
\item{Develop robust and publicly available codes,}
\item{Discover previously unknown degeneracies.}
\end{itemize}

As an example, one could use the data from the microlensing survey
proposed in Section \ref{sec:hband} and inject planets into the light
curves. This would lead to the recovery of the true planetary signals,
and also a characterization of the overall sensitivity of the survey.

There are significant logistical challenges to running such a
competition. However, if it can be undertaken for a modest cost in
time, money, and personnel, the potential benefits to \WF\, and the
field could be substantial.

\section{Conclusions}
\label{sec:conclusions}

\framebox[0.94\textwidth][c]{
\begin{minipage}{0.9\textwidth}
The programs explored by this SAG fall into the following major categories:
\begin{enumerate}
\item{Programs that directly support \WF\, science and reduce its scientific risk:
  \begin{itemize}
    \item{Early, optical, \HST\, imaging of the \WF\, field}
    \item{A preparatory, ground-based, microlensing survey in the near-IR}
  \end{itemize}
}
\item{Programs that develop experience with techniques for measuring planet masses:
  \begin{itemize}
    \item{Satellite parallax observations using \Spitzer, \Kepler, and \TESS}
    \item{\HST\, or AO flux measurements of lenses in ground-based
      microlensing events}
    \item{Measurements of astrometric microlensing for black holes}
  \end{itemize}
}
\item{Programs that support the development of \WF\, analysis pipelines:
  \begin{itemize}
    \item{Multi-epoch \HST/WFC3/IR observations of the bulge}
    \item{An open competition in microlensing analysis}
  \end{itemize}
}
\end{enumerate}
\end{minipage}
}

In our study of programs that will support and enhance the \WF\,
microlensing mission, the program that is the most time critical and
with direct relevance to the \WF\, mission is {\bf optical \HST\,
  precursor imaging of the proposed \WF\,
  fields}. The long time baseline between now and the \WF\, mission is
crucial for measuring proper motions and separately resolving the
source and lens stars, which will necessarily be superposed during the
\WF\, microlensing events. Measured relative proper motions using
\HST\, are the best way to validate the astrometric microlensing
measurements, which will be made by \WF\, and used to measure star and
planet masses for faint lenses. Furthermore, these observations will
not only inform \WF\, field selection by measuring the luminosity
function in the \WF\, fields to faint magnitudes, but will also
provide valuable data for characterizing the source and lens stars by
providing additional colors that can be used as metallicity, effective
temperature, and age diagnostics. These data will vastly enhance the
\WF\, microlensing survey by enabling detailed studies of Galactic
structure using precision astrometry from \WF\,.

The second program with direct relevance to the \WF\, microlensing
mission is {\bf a ground-based near-IR microlensing survey}. Such a
survey would allow direct measurements of the near-IR microlensing
event rates and source luminosity function in the \WF\,
fields. Additionally, the near-IR data from this
survey will provide concurrent observations of microlensing events
where the field overlaps with the optical ground-based surveys. These
data are important for adaptive optics or space-based
luminosity measurements for the lens stars in those events.

In addition to these programs that directly support the \WF\,
microlensing mission, we have identified a series of programs that
will develop expertise and experience in the techniques that \WF\,
will use to measure lens masses and
  locations. {\bf Microlensing parallax observations by \Spitzer\, and
\Kepler\,} would be the first systematic observations for microlens
parallax using a satellite and would serve as pathfinders for a
potential, small, dedicated microlensing parallax satellite. Likewise,
{\bf \HST\, or adaptive optics observations of known microlensing planets}
will measure masses for the host stars and planets by directly
measuring the lens light, a technique that should be routine with
\WF\, but requires additional observations and
  effort for ground-based microlensing. Because all of these programs
will measure masses for microlensing planets, they will also measure
their distances, and hence, will provide the first insight into the
Galactic distribution of planets, which may allow optimization of the
\WF\, fields for maximum planet detection. Similarly, {\bf searches for
astrometric microlensing signals from black holes} will promote
development of a technique that will be frequently used by \WF\, but
is rarely observed today.

Finally, we have identified several other programs of interest to \WF.
{\bf Multi-epoch \HST\, observations} can be used to
  develop and test the mission-critical, \WF\, photometry/astrometry
  pipeline. {\bf A microlensing analysis challenge} will develop
  expertise in and novel approaches to analyzing \WF\, light curves.
Once \WF\, has been selected, we support the enactment of a special
NASA program to fund \WF-related science, including the programs
mentioned here.

\pagebreak

\appendix

\section{Appendix: Charter}
\label{app:charter}

Although the launch of the \WF\, mission is still many years off, it
is nevertheless vitally important to consider what activities must be
carried out in the near future in order to retire any scientific risks
associated with, and maximize the returns from, the \WF\, microlensing
survey. In particular, there may be projects that require a long time
baseline and/or might affect the final mission design, and thus must
be undertaken soon.  This SAG will bring together members of the
microlensing community to identify scientific programs that will
benefit the \WF\, microlensing mission. Of particular interest are
mission-critical observational programs that must be completed before
the launch of \WF.  Specifically, the major question this SAG will
address is:

``What scientific programs can be undertaken now to ensure the success
of the \WF\, mission and maximize its scientific return?"

In the process of answering this question, the SAG will:
\begin{enumerate}
\item{Identify both mission critical and mission enhancing programs,}
\item{Identify immediate science to come out of each program, as well as
the program's direct impact on the \WF\, mission,}
\item{For each proposed program, quantify the improved scientific return
for the \WF\, mission,}
\item{Emphasize programs that can be executed using existing (NASA)
resources.}
\end{enumerate}

\pagebreak

\section{Appendix: Introduction to \WF\, Microlensing}
\label{app:intro}

\subsection{Basic Microlensing Parameters}
\label{app:basics}

\begin{figure}[ht!]
\fbox{
\begin{minipage}{0.6\textwidth}
  \includegraphics[width=0.95\textwidth]{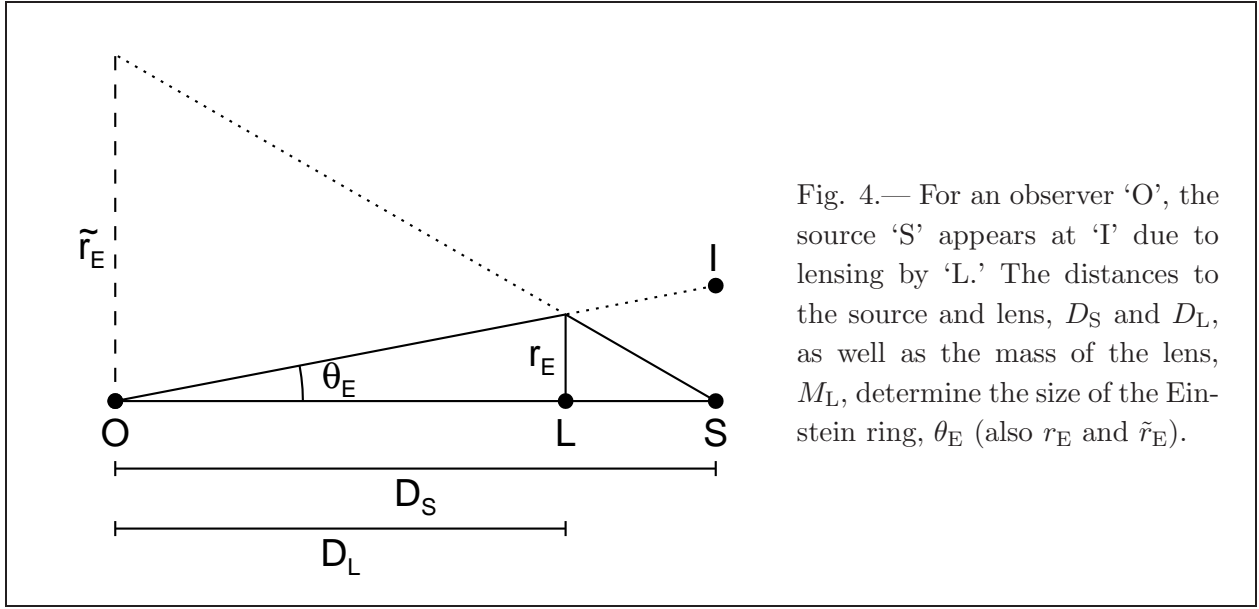}
\end{minipage}
\begin{minipage}{0.34\textwidth}
\caption{For an observer `O', the source `S' appears at `I' due to
  lensing by `L.' The distances to the source and lens, $D_{\rm S}$
  and $D_{\rm L}$, as well as the mass of the lens,
  $M_{\rm L}$, determine the size of the Einstein
  ring, $\thetaE$ (also $r_{\rm E}$ and $\tilde{r}_{\rm
    E}$). \label{fig:geom}}
\end{minipage}
}
\end{figure}

\fbox{\parbox[c][][c]{0.9\textwidth}{\bf The most important scale in
    microlensing is the size of the Einstein ring, $\thetaE$,:
\begin{equation}
\label{eqn:thetaE}
\thetaE = \sqrt{\kappa M_{\rm L} \pi_{\rm rel}}
\quad \mathrm{where} \quad
\pi_{\rm rel} = \frac{\mathrm{AU}}{D_{\rm L}} - \frac{\mathrm{AU}}{D_{\rm S}}
\quad \mathrm{and} \quad
\kappa = 8.14 \mathrm{\, mas\,} M_{\odot}^{-1}.
\end{equation}
$M_{\rm L}$, $D_{\rm L}$, and $D_{\rm S}$ are defined
  as in Figure \ref{fig:geom}.  }}

Planets are detected in microlensing light curves when one of the
images of the source passes over or near the position of the
planet\footnote{A more comprehensive review of
    microlensing can be found in \citet{Gaudi12}.}. Since the
position of those images is determined by $\thetaE$, the inferred
parameters of the planet are necessarily also measured relative to
$\thetaE$. Specifically, the observables are $q=m_{\rm p}/M_{\rm L}$,
the mass of the planet relative to the mass of the lens
star\footnote{i.e., the size of the planet's Einstein
  ring relative to the size of the star's Einstein ring.}, and
$s=r_{\rm proj}/(\thetaE D_{\rm L})$, the position of
the planet projected onto the lens plane relative to the size of the
Einstein ring. Recovering the underlying properties
of the planet requires the measurement of additional
parameters.

\begin{wrapfigure}{r}{0.5\textwidth}
\vspace{-10pt}
\fbox{
\begin{minipage}{0.4\textwidth}
  \bf All of the methods presented here for measuring
    $M_{\rm L}$ require that $\thetaE$ also be measured. Hence,
    whenever $M_{\rm L}$ is measured, so is $D_{\rm L}$ (Equation
    \ref{eqn:thetaE}).
\end{minipage}}
\vspace{-10pt}
\end{wrapfigure}

The direct observable from the light curve is the source magnification
as a function of time, $A(t)$. This magnification is set by the
position of the source relative to the lens, $u$, measured as a
fraction of the Einstein radius, i.e., $A(t) \rightarrow A(u)$. Since
the motion of the source in the lens plane is determined by the
lens-source relative proper motion, $\mu_{\rm rel}$, the value of $u$
depends on this and the size of the Einstein ring, i.e.:
\begin{equation}
u = \sqrt{u_0^2+\tau^2} \quad \mathrm{where} \quad \tau =
\frac{t-t_0}{\tE},
\end{equation}
$u_0$ is the impact parameter between the source and the lens,
$u=u_0$ at $t=t_0$, and
\begin{equation}
\label{eqn:te}
\tE=\frac{\thetaE}{\mu_{\rm rel}}. 
\end{equation}
Although the basic microlensing light curve yields three observable
parameters: $t_0$, $u_0$, and $t_{\rm E}$, only $t_{\rm E}$ encodes
information about the underlying properties of the lens. Hence, we
have only one equation (Equation \ref{eqn:te}) and three
unknowns\footnote{The source can generally be assumed to be in the
  bulge, so $D_{\rm S}$ is generally well determined.}: $M_{\rm L}$,
$D_{\rm L}$, and $\mu_{\rm rel}$. Generally speaking, this is of
limited utility.

In the following sections, we discuss ways to break this degeneracy
using additional observables so that we can measure $M_{\rm L}$ and
$D_{\rm L}$ and recover the intrinsic properties of the planets
$m_{\rm p}$ and $r_{\rm proj}$. Each of these effects
  is optimized in a different parameter regime and has different
  systematics. Hence, measuring multiple effects for the same events
  provides an important comparison sample for quantifying these
  biases.

\subsection{Finite Source Effects}
\label{app:finitesource}

\fbox{\parbox[c][][c]{0.8\textwidth}{\bf $\thetaE$ can be measured if
    the size of the source is resolved by a caustic.}}

As a consequence of the equations of microlensing,
  certain values of $u$ correspond to $A(u)=\infty$ for a
  theoretically perfect, point source; these locations are called the
``caustics''. For a point lens, the caustic is a single point at the
position of the lens star. For a lens with a companion, the caustic is
a closed curve with zero thickness but enclosing a
finite area. If the source star passes over or very close to a caustic
(a common occurrence if a planetary companion is detected), the fact
that the source is not a perfect point source becomes relevant.
The observed magnification is the integration of the magnification
pattern across the face of the source star. Hence, in practice the
magnification is never infinite and it takes a finite amount of time
for source to cross the caustic, creating a rounded peak in the light
curve \citep[e.g. ][]{Gould09}. By measuring the width of this
peak, $2t_{\star}\sim 2\theta_{\star}/\mu_{\rm rel}$, we can determine
the size of the source relative to the size of the Einstein ring,
\begin{equation}
\rho = t_{\star}/\tE = \theta_{\star}/\thetaE,
\end{equation}
where $\theta_{\star}$ is the angular size of the source,
$R_{\star}/D_{\rm S}$.

The position of the source on the color-magnitude diagram can be
combined with surface brightness relations \citep[e.g. ][]{Kervella04} to
estimate $\theta_{\star}$ \citep{Yoo04}. Thus, observing the effects
of the finite size of the source in the light curve yields a
measurement of $\thetaE$ and consequently, $\mu_{\rm rel}$ (Equation
\ref{eqn:te}). {\bf Measuring $\thetaE$ therefore provides a mass-distance
relationship for the lens}, but does not completely resolve the
degeneracy.

\subsection{Microlens Parallax}
\label{app:par}

\begin{figure}[p]
\fbox{
\begin{minipage}[c][][c]{0.94\textwidth}
\centering
\includegraphics[height=0.75\textheight]{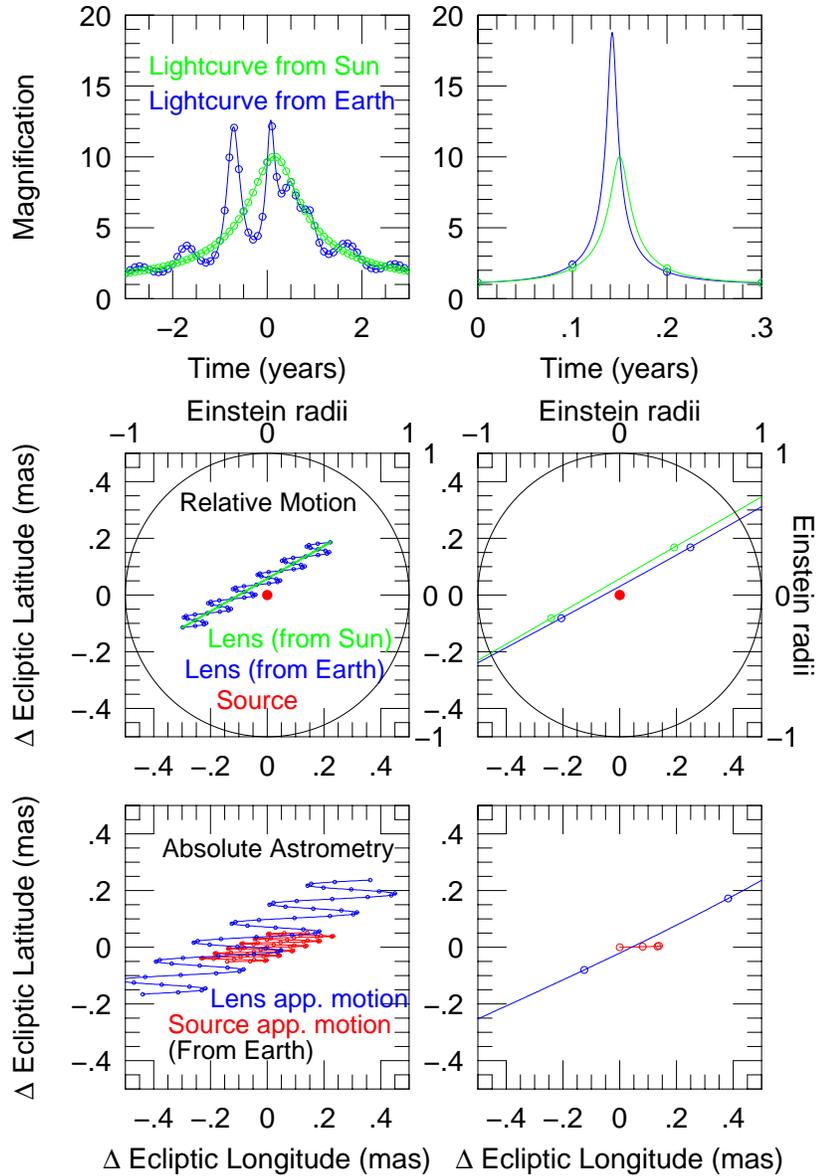}
\caption{Figure 1 from \citet{GouldHorne13}. The orbital parallax
  effect for illustrative (left) and realistic (right) microlensing
  events. Microlensing events appear different from the Earth and the
  Sun because the Earth is an accelerating platform. Bottom: absolute
  trigonometric parallax and proper motion (ppm). Middle: relative
  trigonometric (lower/left labels) and microlensing (upper/right
  labels) ppm. Top: resulting light curves as seen from the Earth
  (blue) and the Sun (green). The same effect will be measurable from
  \WF\, as it orbits the Sun. \label{fig:par}}
\end{minipage}
}
\end{figure}

\fbox{\parbox[c][][c]{0.9\textwidth}{\bf Changing or multiple lines of
    sight lead to parallax effects in the light curve.}}

Because the microlensing effect depends precisely on the {\it
  apparent} alignment of the source and lens stars, slight differences
in the line-of-sight can affect the observed light curve. This
parallax effect can be seen in two different types of
circumstances. First, if there are observations from two different
locations, the two observatories will see two different light curves,
e.g. the ``satellite'' parallax effect between the Earth and a
satellite (see Section \ref{sec:parallax}, Figure
\ref{fig:sat-lc}). Second, orbital parallax effects can arise due to
the acceleration of the observatory during the microlensing event
as the observatory progresses through its orbit
(Figure \ref{fig:par}).

Parallax only affects the light curve by the amount it deflects the
position of the lens relative to the size of the Einstein
ring. Therefore, the observable microlensing parameter is the
microlens parallax, $\piE$, which is related to the trigonometric
parallax by
\begin{equation}
\piE = \frac{\pi_{\rm rel}}{\thetaE}.
\end{equation}
Hence, if {\it both} $\piE$ and $\thetaE$ are measured,
they can be input into Equation \ref{eqn:thetaE} and
both $M_{\rm L}$ and $D_{\rm L}$ are measured.

The major complication with microlens parallax is that it is a
vector. The relative motion between the source and the lens is created
by the combination of proper motion and parallax. The {\it magnitude}
of the relative proper motion is the direct microlensing
observable. Its direction is only relevant if parallax effects are
also observed. Since the known direction of the observatory's orbital
motion or physical baseline sets the orientation of the parallax
effect, this specifies the orientation of the microlensing event on
the sky, and the direction of the proper motion. Hence, in
microlensing, we typically consider the relative proper motion to be a
scalar and the parallax to be a vector whose direction is the
direction of the relative proper motion.

The biggest difficulty with measuring parallax, particularly orbital
parallax, is that {\bf one component of the parallax vector is much
  better constrained than the other.} Consider the fact that a typical
microlensing event lasts much less than a year, during which Earth's
(or \WF's) acceleration is effectively constant. The component of the
parallax parallel to this direction is readily measured because it
accelerates the timing of the light curve producing an asymmetry. In
contrast, the perpendicular component of the parallax produces a
symmetric distortion which can be partially compensated for by other
symmetric parameters of the light curve, e.g. $u_0$ and
$\tE$. Therefore, the parallel component is much more easily measured
than the perpendicular component, so the parallax information can be
incomplete. In fact, \WF\, is expected to measure a large number of
one-dimensional parallaxes for its events, which will provide a
statistical understanding of the underlying lens population
\citep{HanGould95}, but will not give complete parallax measurements
without additional information about the direction of the relative
proper motion. This problem is particularly pronounced
  for objects in the Galactic bulge where the measurement
  uncertainties will be similar to the magnitude of the effect
  \citep{Gould13}.

\subsection{Astrometric Microlensing}
\label{app:ast}

\fbox{\parbox[c][][c]{0.9\textwidth}{\bf The unequal size of the
    lensed images leads to a net shift in the centroid of the light
    that changes over the course of the microlensing event.}}

Figure \ref{fig:ast} illustrates the effect of astrometric
microlensing \citep{Walker95}. For a point lens, the centroid of the
source light appears to execute an ellipse. The expected position of
the centroid relative to the size of the Einstein ring can be
calculated exactly from the microlensing parameters $t_0$, $u_0$, and
$\tE$. Hence, the displacement of the observed centroid directly gives
$\thetaE$ and $\mu_{\rm rel}$. In addition, {\bf the direction of the
  motion of the centroid gives the {\it direction} of the lens-source
  relative proper motion, which is crucial for converting
  one-dimensional parallax measurements into complete measurements of
  $\piE$.}

This method provides an important complement to measuring $\thetaE$
using finite source effects, which requires a caustic
interaction. Such interactions are rare for point lens events; with
mass measurements, these planet-less objects can
provide an important comparison sample for the
planetary microlensing hosts with measured masses. In addition,
because of its precision photometry, \WF\, has the potential to
measure more subtle planetary signals that do not require caustic
crossings \citep{Zhu14}. For these events, astrometric microlensing can be
used to measure $\thetaE$.

\subsection{Measuring Lens Flux}
\label{app:flux}

\fbox{\parbox[c][][c]{0.9\textwidth}{\bf Combining the mass-distance
    relation from Equation \ref{eqn:thetaE} with a
    luminosity-distance relation for the lens allows
    the lens mass to be determined.}}

If the lens star is luminous, it is possible to detect its light in
the presence of the source star. This may be done either when the lens
is superposed or once it has separated slightly from the source. The
main complication with doing this for ground-based microlensing events
is that the seeing-limited PSF may contain light from several
unrelated stars due to the density of stars in the Bulge. However,
with \WF's sensitivity and resolution, which can detect and resolve
main sequence stars in or near the bulge, the lens light can be
measured directly. A measurement of the apparent magnitude of the lens
combined with an extinction relation gives a relationship between the
absolute magnitude of the lens and its distance. The mass-distance
relationship given by $\thetaE$ can be transformed into a second
luminosity-distance relationship using stellar isochrones. Since these
have different functional forms, their intersection determines the
mass and distance to the lens.

The major complication of this method is the possibility of light
contamination. Usually, the most likely origin for the excess light is
the lens star, but the light could also come from a luminous companion
to the lens or the source or, much less likely, an unrelated star that
is still within the confusion limit for diffraction-limited
images. The probabilities for all these scenarios can be quantified
using non-detections from the microlensing light curve, knowledge
about binary star fractions, and the average stellar density in the
field. In addition, if some time has passed since the microlensing
event, it is possible to measure the lens-source relative proper
motion even if the two stars have not completely
separated so as to be clearly resolved. This measurement can be
checked against the value of $\mu_{\rm rel}$ from the light curve to
confirm that the observed light is associated with the lens.

There are two methods that can be used for measuring the relative
proper motion between the source and the putative lens star before
they are clearly resolved.  The first is the color-dependent centroid
shift. If the lens is of comparable brightness to the source but has a
different color, the centroid of the combined lens+source will be
different when observed with different filters. \HST\, follow-up
images of the first two planetary microlensing systems have used this
method to identify the lens stars for these events \citep{Bennett06,
  Dong09_071}. The second method is the elongation of the combined
lens+source image as discussed by \citet{Bennett07}. Both of these
methods require a stable point-spread-function (PSF) that can be
obtained from a space telescope such as \WF. However, these two
methods have only been applied in the few cases referenced here. As in
the case of astrometric microlensing, the best way to validate these
measurements and check for systematics is by comparison to proper
motions measured from precursor \HST\, imaging (Section
\ref{sec:HST}).

\WF\, is expected to directly detect the lens star and its relative
proper motion with respect to the source for more than 75\% of
planetary microlensing events in which the planetary
host star is on the main sequence.

\pagebreak

\section{Appendix: Finding Habitable Zone Planets with \WF}
\label{app:hz}

\fbox{\parbox[c][][c]{0.9\textwidth}{\bf There is a possibility of
    increasing the number of habitable zone planets
    detected by \WF\, by biasing the survey toward
    fields with small $r_{\rm E}$. However, the trade off is likely a
    substantial drop in the total number of planets detected.}}

The nominal \WF\, microlensing survey is expected to find a small
number of ${\sim}$Earth-mass planets in the habitable zones of F-, G-
and K-dwarfs, and a significant number of similar planets just outside
their habitable zones \citep[][in prep]{Penny14}. 

The semimajor axis range that is probed for planets in a given
microlensing event is set by the Einstein radius of the lens
(Appendix \ref{app:basics}). The smaller the Einstein
ring, the smaller the semimajor axis probed. Most habitable planets
will lie inside the Einstein ring, where the detection efficiency
scales approximately as $(a/r_{\mathrm{E}})^3$. In addition to lens
mass, the physical Einstein ring radius depends on
the distances to both the lens and the source 
\begin{equation}
r_{\mathrm{E}} = D_{\rm L}\thetaE = D_{\rm L}\sqrt{\kappa M_{\rm L} \pi_{\rm rel}}.
\end{equation}
Therefore, for a lens of fixed mass, the physical
Einstein radius will be smaller if the lens is either close to the
source or close to the observer.

It is likely possible to find fields with an Einstein radius
distribution skewed towards lower values relative to the fields
optimized for the overall planet detection. For example, sight lines
that exhibit counter moving stellar streams, such as those due to
X-shaped structures in the bulge, \citep{Poleski13}, may provide the
necessary conditions for this. However, {\bf our knowledge of both the
  kinematic and line-of-sight structure of the inner bulge is
  insufficient to predict the locations of such fields at present}. A
two-epoch, two-color survey of the inner bulge with \HST\, would
measure the proper motion and line-of-sight distributions of the F, G,
and K dwarf star populations, including the best
  candidates to be the hosts of habitable planets that are detectable
by microlensing. This would allow the Einstein radius
  distribution for these stars, weighted by the microlensing event
  rate, to be robustly estimated and normalized, allowing a precise
estimate and optimization of the effective number of habitable zones
that would be probed by the \WF\, microlensing
survey.

{\it If such a path were pursued, the boost in habitable planet
  detection efficiency gained by smaller Einstein radii would have to
  more than balance the drop in event rate sustained by moving to
  non-optimal fields.}

\pagebreak
\section{\Euclid\, Microlensing}
\label{sec:Euclid}

\fbox{\parbox[c][][c]{0.9\textwidth}{\bf A \Euclid\, microlensing
    mission concurrent with \WF\, is capable of measuring masses for
    individual free-floating planets.}}

\Euclid\, is expected to launch in 2020, in advance of the \WF\,
mission. NASA is participating in this mission by contributing the
near-IR imagers and a delegation of scientists. This spacecraft with
its near-IR imager is well-suited to microlensing observations and
imaging of the bulge \citep{Penny13}. The scheduling of any
microlensing observations will be likely be decided in 2015-2016. If a
\Euclid\, microlensing survey occurs, it would likely take place
around 2024-2025, which is close to the time of the first \WF\,
microlensing observations.

\subsection{\Euclid-\WF\, Mass Measurements of Free-Floating Planets}

If \Euclid\, microlensing observations are simultaneous or ongoing
with the \WF\, microlensing mission, this provides an opportunity to
measure microlens parallax effects between the two satellites, and
hence, more masses for lens stars and their planets. In fact, this
provides the best opportunity for measuring the masses of
free-floating planets. {\bf Simultaneous observations from \Euclid\,
  at L2 and \WF\, in geosynchronous orbit would be able to directly
  measure masses for 15\% of $10 M_{\oplus}$ free-floating planets},
with the percentage of mass measurements increasing for smaller
free-floating planets.

These microlensing observations could be enhanced by adding a third
ground station for downloading \Euclid\, data. With only two ground
stations, the limits on the data rate mean that high-resolution
optical data cannot be downloaded from \Euclid\, with the same cadence
as the near-IR microlensing data. {\it If NASA or ESA were to
  contribute a third ground station, \Euclid\, would be able to use
  both its optical and near-IR imagers for a microlensing survey},
adding additional color information for its microlensing events. Among
other things, such information can be used to confirm that the
candidate free-floating planet events are indeed caused by
microlensing (an achromatic effect) rather than some unknown
astrophysical phenomenon, which would likely have some color
dependence \citep[e.g. ][]{Gouldetal13}. Multi-band data can also be
used to vet anomalous microlensing events whose light curves may be
explained by either a 2-body lens or a binary source. If the proposed
binary source has unequal mass components, the changing, differential
magnification of the two sources will result in chromatic effects in
the light curve \citep{Gaudi98,Hwangetal13}.

\subsection{Early \Euclid\, Imaging of the Bulge}

During its commissioning, \Euclid\, could conduct early imaging of the
Galactic bulge. Deep, multiband (optical, $Y$, $J$, and $H$)
observations could serve as a precursor to the \Euclid\, microlensing
mission.  The optical data will be useful for characterizing the stars
in the \WF\, field. The improved resolution over the near-IR will aid
in proper motion studies of \WF\, stars and could improve \WF\,
photometry because the positions of the stars will be better known. In
addition, the optical luminosity functions will complement the data
from \WF. However, these data will not replace precursor \HST\,
imaging (Section \ref{sec:HST}), because they will not have the time
baseline or resolution to resolve the future sources and lenses, a
requirement for measuring the relative lens-source proper
motions. Hence, although an early epoch of multi-band data from
\Euclid\, may not be of substantial benefit to the \WF\, microlensing
survey, it will certainly be beneficial for \WF\, science in general.

\pagebreak
\bibliographystyle{apj}

\begin{thebibliography}{}

\expandafter\ifx\csname natexlab\endcsname\relax\def\natexlab#1{#1}\fi

\bibitem[{{Batista} {et~al.}(2014){Batista}, {Beaulieu}, {Gould}, {Bennett},
  {Yee}, {Fukui}, {Gaudi}, {Sumi}, \& {Udalski}}]{Batista14}
{Batista}, V., {Beaulieu}, J.-P., {Gould}, A., {et~al.} 2014, \apj, 780, 54

\bibitem[{{Batista} {et~al.} (2014)}]{Batista14b}
{Batista}, V., {et al.} 2014, in prep

\bibitem[{{Bennett}(2010)}]{Bennett10}
{Bennett}, D.~P. 2010, \apj, 716, 1408

\bibitem[{{Bennett} {et~al.}(2006){Bennett}, {Anderson}, {Bond}, {Udalski}, \&
  {Gould}}]{Bennett06}
{Bennett}, D.~P., {Anderson}, J., {Bond}, I.~A., {Udalski}, A., \& {Gould}, A.
  2006, \apjl, 647, L171

\bibitem[{{Bennett} {et~al.}(2007){Bennett}, {Anderson}, \&
  {Gaudi}}]{Bennett07}
{Bennett}, D.~P., {Anderson}, J., \& {Gaudi}, B.~S. 2007, \apj, 660, 781

\bibitem[{{Bennett} \& {Rhie}(2002)}]{BennettRhie02}
{Bennett}, D.~P., \& {Rhie}, S.~H. 2002, \apj, 574, 985

\bibitem[{{Bozza}(2010)}]{Bozza10}
{Bozza}, V. 2010, \mnras, 408, 2188

\bibitem[{{Cassan}(2008)}]{Cassan08}
{Cassan}, A. 2008, \aap, 491, 587

\bibitem[{{Dong} {et~al.}(2006){Dong}, {DePoy}, {Gaudi}, {Gould}, {Han},
  {Park}, {Pogge}, {MuFun Collaboration}, {Udalski}, {Szewczyk}, {Kubiak},
  {Szyma{\'n}ski}, {Pietrzy{\'n}ski}, {Soszy{\'n}ski}, {Wyrzykowski},
  {{\.Z}ebru{\'n}}, \& {OGLE Collaboration}}]{Dong06}
{Dong}, S., {DePoy}, D.~L., {Gaudi}, B.~S., {et~al.} 2006, \apj, 642, 842

\bibitem[{{Dong} {et~al.}(2007){Dong}, {Udalski}, {Gould}, {Reach}, {Christie},
  {Boden}, {Bennett}, {Fazio}, {Griest}, {Szyma{\'n}ski}, {Kubiak},
  {Soszy{\'n}ski}, {Pietrzy{\'n}ski}, {Szewczyk}, {Wyrzykowski}, {Ulaczyk},
  {Wieckowski}, {Paczy{\'n}ski}, {DePoy}, {Pogge}, {Preston}, {Thompson}, \&
  {Patten}}]{Dong07}
{Dong}, S., {Udalski}, A., {Gould}, A., {et~al.} 2007, \apj, 664, 862

\bibitem[{{Dong} {et~al.}(2009){Dong}, {Gould}, {Udalski}, {Anderson},
  {Christie}, {Gaudi}, {The OGLE Collaboration}, {Jaroszy{\'n}ski}, {Kubiak},
  {Szyma{\'n}ski}, {Pietrzy{\'n}ski}, {Soszy{\'n}ski}, {Szewczyk}, {Ulaczyk},
  {Wyrzykowski}, {The {$\mu$}FUN Collaboration}, {DePoy}, {Fox}, {Gal-Yam},
  {Han}, {L{\'e}pine}, {McCormick}, {Ofek}, {Park}, {Pogge}, {The MOA
  Collaboration}, {Abe}, {Bennett}, {Bond}, {Britton}, {Gilmore}, {Hearnshaw},
  {Itow}, {Kamiya}, {Kilmartin}, {Korpela}, {Masuda}, {Matsubara}, {Motomura},
  {Muraki}, {Nakamura}, {Ohnishi}, {Okada}, {Rattenbury}, {Saito}, {Sako},
  {Sasaki}, {Sullivan}, {Sumi}, {Tristram}, {Yanagisawa}, {Yock}, {Yoshoika},
  {The PLANET/Robo Net Collaborations}, {Albrow}, {Beaulieu}, {Brillant},
  {Calitz}, {Cassan}, {Cook}, {Coutures}, {Dieters}, {Prester}, {Donatowicz},
  {Fouqu{\'e}}, {Greenhill}, {Hill}, {Hoffman}, {Horne}, {J{\o}rgensen},
  {Kane}, {Kubas}, {Marquette}, {Martin}, {Meintjes}, {Menzies}, {Pollard},
  {Sahu}, {Vinter}, {Wambsganss}, {Williams}, {Bode}, {Bramich}, {Burgdorf},
  {Snodgrass}, {Steele}, {Doublier}, \& {Foellmi}}]{Dong09_071}
{Dong}, S., {Gould}, A., {Udalski}, A., {et~al.} 2009, \apj, 695, 970

\bibitem[{{Gaudi}(1998)}]{Gaudi98}
{Gaudi}, B.~S. 1998, \apj, 506, 533

\bibitem[{{Gaudi}(2012)}]{Gaudi12}
---. 2012, \araa, 50, 411

\bibitem[{{Gaudi} \& {Gould}(1997)}]{GaudiGould97}
{Gaudi}, B.~S., \& {Gould}, A. 1997, \apj, 486, 85

\bibitem[{{Gaudi} {et~al.}(2008){Gaudi}, {Bennett}, {Udalski}, {Gould},
  {Christie}, {Maoz}, {Dong}, {McCormick}, {Szyma{\'n}ski}, {Tristram},
  {Nikolaev}, {Paczy{\'n}ski}, {Kubiak}, {Pietrzy{\'n}ski}, {Soszy{\'n}ski},
  {Szewczyk}, {Ulaczyk}, {Wyrzykowski}, {OGLE Collaboration}, {DePoy}, {Han},
  {Kaspi}, {Lee}, {Mallia}, {Natusch}, {Pogge}, {Park}, {{$\mu$}-Fun
  Collabortion}, {Abe}, {Bond}, {Botzler}, {Fukui}, {Hearnshaw}, {Itow},
  {Kamiya}, {Korpela}, {Kilmartin}, {Lin}, {Masuda}, {Matsubara}, {Motomura},
  {Muraki}, {Nakamura}, {Okumura}, {Ohnishi}, {Rattenbury}, {Sako}, {Saito},
  {Sato}, {Skuljan}, {Sullivan}, {Sumi}, {Sweatman}, {Yock}, {MOA
  Collaboration}, {Albrow}, {Allan}, {Beaulieu}, {Burgdorf}, {Cook},
  {Coutures}, {Dominik}, {Dieters}, {Fouqu{\'e}}, {Greenhill}, {Horne},
  {Steele}, {Tsapras}, {Planet Collaboration}, {RoboNet Collaborations},
  {Chaboyer}, {Crocker}, {Frank}, \& {Macintosh}}]{Gaudi08}
{Gaudi}, B.~S., {Bennett}, D.~P., {Udalski}, A., {et~al.} 2008, Science, 319,
  927

\bibitem[{{Gonzalez} {et~al.}(2012){Gonzalez}, {Rejkuba}, {Zoccali}, {Valenti},
  {Minniti}, {Schultheis}, {Tobar}, \& {Chen}}]{Gonzalezetal12}
{Gonzalez}, O.~A., {Rejkuba}, M., {Zoccali}, M., {et~al.} 2012, \aap, 543, A13

\bibitem[{{Gould}(1995)}]{Gould95}
{Gould}, A. 1995, \apjl, 446, L71

\bibitem[{{Gould}(2013)}]{Gould13}
---. 2013, \apjl, 763, L35

\bibitem[{{Gould} {et al.}(2014)}]{Gouldetal14}
{Gould}, A., {et al.} 2014, Science, submitted

\bibitem[{{Gould} \& {Gaucherel}(1997)}]{GouldGaucherel97}
{Gould}, A., \& {Gaucherel}, C. 1997, \apj, 477, 580

\bibitem[{{Gould} \& {Horne}(2013)}]{GouldHorne13}
{Gould}, A., \& {Horne}, K. 2013, \apjl, 779, L28

\bibitem[{{Gould} \& {Loeb}(1992)}]{GouldLoeb92}
{Gould}, A., \& {Loeb}, A. 1992, \apj, 396, 104

\bibitem[{{Gould} {et~al.}(2009){Gould}, {Udalski}, {Monard}, {Horne}, {Dong},
  {Miyake}, {Sahu}, {Bennett}, {Wyrzykowski}, {Soszy{\'n}ski}, {Szyma{\'n}ski},
  {Kubiak}, {Pietrzy{\'n}ski}, {Szewczyk}, {Ulaczyk}, {OGLE Collaboration},
  {Allen}, {Christie}, {DePoy}, {Gaudi}, {Han}, {Lee}, {McCormick}, {Natusch},
  {Park}, {Pogge}, {{$\mu$}FUN Collaboration}, {Allan}, {Bode}, {Bramich},
  {Burgdorf}, {Dominik}, {Fraser}, {Kerins}, {Mottram}, {Snodgrass}, {Steele},
  {Street}, {Tsapras}, {RoboNet Collaboration}, {Abe}, {Bond}, {Botzler},
  {Fukui}, {Furusawa}, {Hearnshaw}, {Itow}, {Kamiya}, {Kilmartin}, {Korpela},
  {Lin}, {Ling}, {Masuda}, {Matsubara}, {Muraki}, {Nagaya}, {Ohnishi},
  {Okumura}, {Perrott}, {Rattenbury}, {Saito}, {Sako}, {Skuljan}, {Sullivan},
  {Sumi}, {Sweatman}, {Tristram}, {Yock}, {MOA Collaboration}, {Albrow},
  {Beaulieu}, {Coutures}, {Calitz}, {Caldwell}, {Fouque}, {Martin}, {Williams},
  \& {PLANET Collaboration}}]{Gould09}
{Gould}, A., {Udalski}, A., {Monard}, B., {et~al.} 2009, \apjl, 698, L147

\bibitem[{{Gould} {et~al.}(2013){Gould}, {Yee}, {Bond}, {Udalski}, {Han},
  {J{\o}rgensen}, {Greenhill}, {Tsapras}, {Pinsonneault}, {Bensby}, {Allen},
  {Almeida}, {Bos}, {Christie}, {DePoy}, {Dong}, {Gaudi}, {Hung}, {Jablonski},
  {Lee}, {McCormick}, {Moorhouse}, {Mu{\~n}oz}, {Natusch}, {Nola}, {Pogge},
  {Skowron}, {Thornley}, {The {$\mu$}FUN Collaboration}, {Abe}, {Bennett},
  {Botzler}, {Chote}, {Freeman}, {Fukui}, {Furusawa}, {Harris}, {Itow}, {Ling},
  {Masuda}, {Matsubara}, {Miyake}, {Ohnishi}, {Rattenbury}, {Saito},
  {Sullivan}, {Sumi}, {Suzuki}, {Sweatman}, {Tristram}, {Wada}, {Yock}, {The
  MOA Collaboration}, {Szyma{\'n}ski}, {Soszy{\'n}ski}, {Kubiak}, {Poleski},
  {Ulaczyk}, {Pietrzy{\'n}ski}, {Wyrzykowski}, {The OGLE Collaboration},
  {Alsubai}, {Bozza}, {Browne}, {Burgdorf}, {Calchi Novati}, {Dodds},
  {Dominik}, {Finet}, {Gerner}, {Hardis}, {Harps{\o}e}, {Hessman}, {Hinse},
  {Hundertmark}, {Kains}, {Kerins}, {Liebig}, {Mancini}, {Mathiasen}, {Penny},
  {Proft}, {Rahvar}, {Ricci}, {Sahu}, {Scarpetta}, {Sch{\"a}fer},
  {Sch{\"o}nebeck}, {Snodgrass}, {Southworth}, {Surdej}, {Wambsganss},
  {MiNDSTEp Consortium}, {Street}, {Horne}, {Bramich}, {Steele}, {The RoboNet
  Collaboration}, {Albrow}, {Bachelet}, {Batista}, {Beatty}, {Beaulieu},
  {Bennett}, {Bowens-Rubin}, {Brillant}, {Caldwell}, {Cassan}, {Cole},
  {Corrales}, {Coutures}, {Dieters}, {Dominis Prester}, {Donatowicz},
  {Fouqu{\'e}}, {Henderson}, {Kubas}, {Marquette}, {Martin}, {Menzies},
  {Shappee}, {Williams}, {van Saders}, {Zub}, \& {The PLANET
  Collaboration}}]{Gouldetal13}
{Gould}, A., {Yee}, J.~C., {Bond}, I.~A., {et al.} 2013, \apj, 763, 141

\bibitem[{{Gould} \& {Yee}(2014)}]{GouldYee14}
{Gould}, A., \& {Yee}, J.~C. 2014, \apj, 784, 64

\bibitem[{{Green} {et~al.}(2012){Green}, {Schechter}, {Baltay}, {Bean},
  {Bennett}, {Brown}, {Conselice}, {Donahue}, {Fan}, {Gaudi}, {Hirata},
  {Kalirai}, {Lauer}, {Nichol}, {Padmanabhan}, {Perlmutter}, {Rauscher},
  {Rhodes}, {Roellig}, {Stern}, {Sumi}, {Tanner}, {Wang}, {Weinberg}, {Wright},
  {Gehrels}, {Sambruna}, {Traub}, {Anderson}, {Cook}, {Garnavich},
  {Hillenbrand}, {Ivezic}, {Kerins}, {Lunine}, {McDonald}, {Penny}, {Phillips},
  {Rieke}, {Riess}, {van der Marel}, {Barry}, {Cheng}, {Content}, {Cutri},
  {Goullioud}, {Grady}, {Helou}, {Jackson}, {Kruk}, {Melton}, {Peddie},
  {Rioux}, \& {Seiffert}}]{Green12}
{Green}, J., {Schechter}, P., {Baltay}, C., {et~al.} 2012, ArXiv e-prints,
  arXiv:1208.4012

\bibitem[{{Griest} \& {Safizadeh}(1998)}]{GriestSafizadeh98}
{Griest}, K., \& {Safizadeh}, N. 1998, \apj, 500, 37

\bibitem[{{Han} \& {Gould}(1995)}]{HanGould95}
{Han}, C., \& {Gould}, A. 1995, \apj, 447, 53

\bibitem[{{Han} \& {Gould}(1996)}]{HanGould96}
---. 1996, \apj, 467, 540

\bibitem[{{Hwang} {et~al.}(2013){Hwang}, {Choi}, {Bond}, {Sumi}, {Han},
  {Gaudi}, {Gould}, {Bozza}, {Beaulieu}, {Tsapras}, {Abe}, {Bennett},
  {Botzler}, {Chote}, {Freeman}, {Fukui}, {Fukunaga}, {Harris}, {Itow},
  {Koshimoto}, {Ling}, {Masuda}, {Matsubara}, {Muraki}, {Namba}, {Ohnishi},
  {Rattenbury}, {Saito}, {Sullivan}, {Sweatman}, {Suzuki}, {Tristram}, {Wada},
  {Yamai}, {Yock}, {Yonehara}, {The MOA Collaboration}, {de Almeida}, {DePoy},
  {Dong}, {Jablonski}, {Jung}, {Kavka}, {Lee}, {Park}, {Pogge}, {Shin}, {Yee},
  {The {$\mu$}FUN Collaboration}, {Albrow}, {Bachelet}, {Batista}, {Brillant},
  {Caldwell}, {Cassan}, {Cole}, {Corrales}, {Coutures}, {Dieters}, {Dominis
  Prester}, {Donatowicz}, {Fouqu{\'e}}, {Greenhill}, {J{\o}rgensen}, {Kane},
  {Kubas}, {Marquette}, {Martin}, {Meintjes}, {Menzies}, {Pollard}, {Williams},
  {Wouters}, {The PLANET Collaboration}, {Bramich}, {Dominik}, {Horne},
  {Browne}, {Hundertmark}, {Ipatov}, {Kains}, {Snodgrass}, {Steele}, {Street},
  \& {The RoboNet Collaboration}}]{Hwangetal13}
{Hwang}, K.-H., {Choi}, J.-Y., {Bond}, I.~A., {et~al.} 2013, \apj, 778, 55

\bibitem[{{Janczak} {et~al.}(2010){Janczak}, {Fukui}, {Dong}, {Monard},
  {Koz{\l}owski}, {Gould}, {Beaulieu}, {Kubas}, {Marquette}, {Sumi}, {Bond},
  {Bennett}, {Abe}, {Furusawa}, {Hearnshaw}, {Hosaka}, {Itow}, {Kamiya},
  {Korpela}, {Kilmartin}, {Lin}, {Ling}, {Makita}, {Masuda}, {Matsubara},
  {Miyake}, {Muraki}, {Nagaya}, {Nagayama}, {Nishimoto}, {Ohnishi}, {Perrott},
  {Rattenbury}, {Sako}, {Saito}, {Skuljan}, {Sullivan}, {Sweatman}, {Tristram},
  {Yock}, {The MOA Collaboration}, {An}, {Christie}, {Chung}, {DePoy}, {Gaudi},
  {Han}, {Lee}, {Mallia}, {Natusch}, {Park}, {Pogge}, {The {$\mu$}FUN
  Collaboration}, {Anguita}, {Calchi Novati}, {Dominik}, {J{\o}rgensen},
  {Masi}, {Mathiasen}, {The MiNDSTEp Collaboration}, {Batista}, {Brillant},
  {Cassan}, {Cole}, {Corrales}, {Coutures}, {Dieters}, {Fouqu{\'e}},
  {Greenhill}, \& {The PLANET Collaboration}}]{Janczak10}
{Janczak}, J., {Fukui}, A., {Dong}, S., {et~al.} 2010, \apj, 711, 731

\bibitem[{{Kervella} {et~al.}(2004){Kervella}, {Th{\'e}venin}, {Di Folco}, \&
  {S{\'e}gransan}}]{Kervella04}
{Kervella}, P., {Th{\'e}venin}, F., {Di Folco}, E., \& {S{\'e}gransan}, D.
  2004, \aap, 426, 297

\bibitem[{{Kubas} {et~al.}(2012){Kubas}, {Beaulieu}, {Bennett}, {Cassan},
  {Cole}, {Lunine}, {Marquette}, {Dong}, {Gould}, {Sumi}, {Batista},
  {Fouqu{\'e}}, {Brillant}, {Dieters}, {Coutures}, {Greenhill}, {Bond},
  {Nagayama}, {Udalski}, {Pompei}, {N{\"u}rnberger}, \& {Le Bouquin}}]{Kubas12}
{Kubas}, D., {Beaulieu}, J.~P., {Bennett}, D.~P., {et~al.} 2012, \aap, 540, A78

\bibitem[{{Muraki} {et~al.}(2011){Muraki}, {Han}, {Bennett}, {Suzuki},
  {Monard}, {Street}, {Jorgensen}, {Kundurthy}, {Skowron}, {Becker}, {Albrow},
  {Fouqu{\'e}}, {Heyrovsk{\'y}}, {Barry}, {Beaulieu}, {Wellnitz}, {Bond},
  {Sumi}, {Dong}, {Gaudi}, {Bramich}, {Dominik}, {Abe}, {Botzler}, {Freeman},
  {Fukui}, {Furusawa}, {Hayashi}, {Hearnshaw}, {Hosaka}, {Itow}, {Kamiya},
  {Korpela}, {Kilmartin}, {Lin}, {Ling}, {Makita}, {Masuda}, {Matsubara},
  {Miyake}, {Nishimoto}, {Ohnishi}, {Perrott}, {Rattenbury}, {Saito},
  {Skuljan}, {Sullivan}, {Sweatman}, {Tristram}, {Wada}, {Yock}, {MOA
  Collaboration}, {Christie}, {DePoy}, {Gorbikov}, {Gould}, {Kaspi}, {Lee},
  {Mallia}, {Maoz}, {McCormick}, {Moorhouse}, {Natusch}, {Park}, {Pogge},
  {Polishook}, {Shporer}, {Thornley}, {Yee}, {{$\mu$}FUN Collaboration},
  {Allan}, {Browne}, {Horne}, {Kains}, {Snodgrass}, {Steele}, {Tsapras},
  {RoboNet Collaboration}, {Batista}, {Bennett}, {Brillant}, {Caldwell},
  {Cassan}, {Cole}, {Corrales}, {Coutures}, {Dieters}, {Dominis Prester},
  {Donatowicz}, {Greenhill}, {Kubas}, {Marquette}, {Martin}, {Menzies}, {Sahu},
  {Waldman}, {Williams}, {Zub}, {PLANET Collaboration}, {Bourhrous},
  {Matsuoka}, {Nagayama}, {Oi}, {Randriamanakoto}, {IRSF Observers}, {Bozza},
  {Burgdorf}, {Calchi Novati}, {Dreizler}, {Finet}, {Glitrup}, {Harps{\o}e},
  {Hinse}, {Hundertmark}, {Liebig}, {Maier}, {Mancini}, {Mathiasen}, {Rahvar},
  {Ricci}, {Scarpetta}, {Skottfelt}, {Surdej}, {Southworth}, {Wambsganss},
  {Zimmer}, {MiNDSTEp Consortium}, {Udalski}, {Poleski}, {Wyrzykowski},
  {Ulaczyk}, {Szyma{\'n}ski}, {Kubiak}, {Pietrzy{\'n}ski}, {Soszy{\'n}ski}, \&
  {OGLE Collaboration}}]{Muraki11}
{Muraki}, Y., {Han}, C., {Bennett}, D.~P., {et~al.} 2011, \apj, 741, 22

\bibitem[{{Penny} {et~al.}(2013) {Penny}, {Kerins}, {Rattenbury},
  {Beaulieu}, {Robin}, {Mao}, {Batista}, {Calchi Novati}, {Cassan},
  {Fouqu{\'e}}, {McDonald}, {Marquette}, {Tisserand}, \& {Zapatero
    Osorio}}]{Penny13}
{Penny}, M.~T., {Kerins}, E., {Rattenbury}, N., {et~al.} 2013, \mnras, 434, 2

\bibitem[{{Penny} {et~al.} (2014)}]{Penny14}
{Penny}, M., {et~al.} 2014, in prep

\bibitem[{{Poleski} {et~al.}(2013){Poleski}, {Udalski}, {Gould},
  {Szyma{\'n}ski}, {Soszy{\'n}ski}, {Kubiak}, {Pietrzy{\'n}ski}, {Ulaczyk}, \&
  {Wyrzykowski}}]{Poleski13}
{Poleski}, R., {Udalski}, A., {Gould}, A., {et~al.} 2013, \apj, 776, 76

\bibitem[{{Robin} {et~al.}(2003){Robin}, {Reyl{\'e}}, {Derri{\`e}re}, \&
  {Picaud}}]{Robinetal03}
{Robin}, A.~C., {Reyl{\'e}}, C., {Derri{\`e}re}, S., \& {Picaud}, S. 2003,
  \aap, 409, 523

\bibitem[{{Sahu} {et~al.}(2012){Sahu}, {Albrow}, {Anderson}, {Bond}, {Bond},
  {Brown}, {Casertano}, {Dominik}, {Ferguson}, {Fryer}, {Livio}, {Mao},
  {Perrott}, {Udalski}, \& {Yock}}]{Sahu12}
{Sahu}, K.~C., {Albrow}, M., {Anderson}, J., {et~al.} 2012, in American
  Astronomical Society Meeting Abstracts, Vol. 220, American Astronomical
  Society Meeting Abstracts \#220, \#307.03

\bibitem[{{Spergel} {et~al.}(2013){Spergel}, {Gehrels}, {Breckinridge},
  {Donahue}, {Dressler}, {Gaudi}, {Greene}, {Guyon}, {Hirata}, {Kalirai},
  {Kasdin}, {Moos}, {Perlmutter}, {Postman}, {Rauscher}, {Rhodes}, {Wang},
  {Weinberg}, {Centrella}, {Traub}, {Baltay}, {Colbert}, {Bennett},
  {Kiessling}, {Macintosh}, {Merten}, {Mortonson}, {Penny}, {Rozo},
  {Savransky}, {Stapelfeldt}, {Zu}, {Baker}, {Cheng}, {Content}, {Dooley},
  {Foote}, {Goullioud}, {Grady}, {Jackson}, {Kruk}, {Levine}, {Melton},
  {Peddie}, {Ruffa}, \& {Shaklan}}]{Spergel13}
{Spergel}, D., {Gehrels}, N., {Breckinridge}, J., {et~al.} 2013, ArXiv
  e-prints, arXiv:1305.5422

\bibitem[{{Sumi} {et~al.}(2011){Sumi}, {Kamiya}, {Bennett}, {Bond}, {Abe},
  {Botzler}, {Fukui}, {Furusawa}, {Hearnshaw}, {Itow}, {Kilmartin}, {Korpela},
  {Lin}, {Ling}, {Masuda}, {Matsubara}, {Miyake}, {Motomura}, {Muraki},
  {Nagaya}, {Nakamura}, {Ohnishi}, {Okumura}, {Perrott}, {Rattenbury}, {Saito},
  {Sako}, {Sullivan}, {Sweatman}, {Tristram}, {Udalski}, {Szyma{\'n}ski},
  {Kubiak}, {Pietrzy{\'n}ski}, {Poleski}, {Soszy{\'n}ski}, {Wyrzykowski},
  {Ulaczyk}, \& {Microlensing Observations in Astrophysics (MOA)
  Collaboration}}]{Sumietal11}
{Sumi}, T., {Kamiya}, K., {Bennett}, D.~P., {et~al.} 2011, \nat, 473, 349

\bibitem[{{Sumi} {et~al.}(2013){Sumi}, {Bennett}, {Bond}, {Abe}, {Botzler},
  {Fukui}, {Furusawa}, {Itow}, {Ling}, {Masuda}, {Matsubara}, {Muraki},
  {Ohnishi}, {Rattenbury}, {Saito}, {Sullivan}, {Suzuki}, {Sweatman},
  {Tristram}, {Wada}, {Yock}, \& {MOA Collaboratoin}}]{Sumietal13}
{Sumi}, T., {Bennett}, D.~P., {Bond}, I.~A., {et~al.} 2013, \apj, 778, 150

\bibitem[{{Walker}(1995)}]{Walker95}
{Walker}, M.~A. 1995, \apj, 453, 37

\bibitem[{{Yee}(2013)}]{Yee13}
{Yee}, J.~C. 2013, \apjl, 770, L31

\bibitem[{{Yoo} {et~al.}(2004){Yoo}, {DePoy}, {Gal-Yam}, {Gaudi}, {Gould},
  {Han}, {Lipkin}, {Maoz}, {Ofek}, {Park}, {Pogge}, {Mu-Fun Collaboration},
  {Udalski}, {Soszy{\'n}ski}, {Wyrzykowski}, {Kubiak}, {Szyma{\'n}ski},
  {Pietrzy{\'n}ski}, {Szewczyk}, {{\.Z}ebru{\'n}}, \& {OGLE
  Collaboration}}]{Yoo04}
{Yoo}, J., {DePoy}, D.~L., {Gal-Yam}, A., {et~al.} 2004, \apj, 603, 139

\bibitem[{{Zhu} {et al.}(2014){Zhu}, {Penny}, {Mao}, {Gould}, \&
    {Gendron}}]{Zhu14}
{Zhu}, W. and {Penny}, M. and {Mao}, S., {et~al.} 2014, \apj, 788, 73

\end{thebibliography}

\end{document}